\documentclass[a4paper,11pt]{article}
\usepackage{jinstpub} 
\usepackage{lineno}
\usepackage{float} 
\usepackage{amsmath}
\usepackage{graphicx} 
\usepackage{subcaption}
\usepackage{comment}

\title{\boldmath Design and Simulation Study of the Hadronic Calorimeter for the EicC Zero Degree Calorimeter}

\author[a]{Maiyu Wang,}
\author[a]{Yuan Li,}
\author[a]{Hao Wang,}
\author[a]{Xikun Sun,}
\author[a]{Zixiang Zhou,}
\author[b,c,d]{Yutie Liang,}
\author[a]{Weizhi Xiong,}
\author[b,1]{Ye Tian,\note{Corresponding author.}}
\author[a,2]{Ting Lin\note{Corresponding author.}}

\affiliation[a]{Key Laboratory of Particle Physics and Particle Irradiation (MOE), Institute of Frontier and Interdisciplinary Science, Shandong University, Qingdao, Shandong 266237, China}
\affiliation[b]{Institute of Modern Physics, Chinese Academy of Sciences, Lanzhou, Gansu Province 730000, China}
\affiliation[c]{School of Nuclear Science and Technology, University of Chinese Academy of Sciences, Beijing 100049, China}
\affiliation[d]{Heavy Ion Science and Technology Key Laboratory, Institute of Modern Physics, Chinese Academy of Sciences, Lanzhou 730000, China}

\emailAdd{tianye@impcas.ac.cn}
\emailAdd{tinglin@sdu.edu.cn}

\abstract{The Zero-Degree Calorimeter (ZDC) at the proposed Electron-Ion Collider in China (EicC) is essential for detecting forward-going neutral particles and supporting the core nucleon spin and 3D imaging physics programs. In this work, a highly compact Spaghetti Calorimeter (SPACAL) architecture is proposed and optimized as the baseline design for the ZDC hadronic section. To systematically and quantitatively evaluate its physics capabilities, a comprehensive end-to-end Geant4 simulation framework was developed, integrating energy deposition, optical photon transport, photo-detection, and modeled front-end signal digitization. The optimized detector demonstrates excellent performance for neutron detection, achieving an energy resolution of $33.42\%/\sqrt{E/\mathrm{GeV}} + 1.47\%$ and a timing resolution of approximately 500 ps, outperforming the targeted specification. Furthermore, full-system simulations incorporating the upstream electromagnetic calorimeter yield a sub-centimeter transverse position resolution. Moreover, topological shower shape analysis across the full detector system provides robust particle identification (PID), with photon-neutron separation accuracy reaching over $99\%$ for high-energy incident particles. These quantitative results confirm that the SPACAL design fully satisfies the stringent operational and physics requirements of the EicC forward kinematic region.
}


\begin{document}
\maketitle
\flushbottom

\section{Introduction}
\label{sec:intro}

The proposed Electron-Ion Collider in China (EicC) is a next-generation high-luminosity facility designed to probe the internal structure of nucleons and nuclei, with a particular focus on the sea quark kinematic region~\cite{Anderle:2021wcy}. One of the primary goals of the EicC physics program is to achieve high-precision three-dimensional imaging of the nucleon through Generalized Parton Distributions (GPDs)~\cite{Ji:2004gf,Chakrabarti:2005zm,Scopetta:2003et,Burkert:2022hjz,Mezrag:2023nkp,Qiu:2024pvw} and Transverse Momentum Dependent (TMD) parton distributions~\cite{Boussarie:2023izj, Angeles-Martinez:2015sea}, while also addressing broader questions in hadron structure, including the emergence of hadron mass and the partonic structure of light mesons~\cite{Roberts:2021nhw,Arrington:2021biu,Ding:2022ows,Yao:2024drm}. Furthermore, the facility provides a unique environment to probe the production mechanisms of exotic hadrons—such as $XYZ$ states and hidden-charm pentaquarks—via near-threshold photoproduction and electroproduction~\cite{Chen:2016qju, Guo:2017jvc,Brambilla:2019esw}. To accomplish these ambitious physics goals, a comprehensive and highly performant suite of detectors is imperative, particularly in the extreme forward rapidity regions where critical kinematic information is concentrated.

In deep inelastic scattering (DIS) and a variety of exclusive and diffractive processes, crucial momentum and energy fractions are carried by final-state particles emitted at near-zero angles relative to the incident ion beam. Capturing these forward-going particles is essential for fully reconstructing event kinematics and guaranteeing event exclusivity. For instance, accessing GPDs via deeply virtual meson production (DVMP, e.g., $ep \to en\pi^+$)~\cite{Favart:2015umi} requires the precise tagging of forward recoil nucleons. Similarly, investigating the structure functions and form factors of mesons, such as pions and kaons, relies heavily on the Sullivan process~\cite{Sullivan:1971kd, Aguilar:2019teb, Lu:2025bnm,Paul:2025lambda}, where the electron scatters off the virtual meson cloud surrounding the proton; isolating this kinematic regime requires the explicit detection of ultra-forward spectator neutrons. Furthermore, in electron-nucleus ($e$-$A$) collisions, spectator-neutron detection plays a critical role in determining collision centrality and understanding nuclear fragmentation dynamics~\cite{Hegazy:2024tgt,AbdulKhalek:2021gbh,Zheng:2014cha}.

\begin{figure}[htbp]
    \centering
    \includegraphics[width=0.95\textwidth]{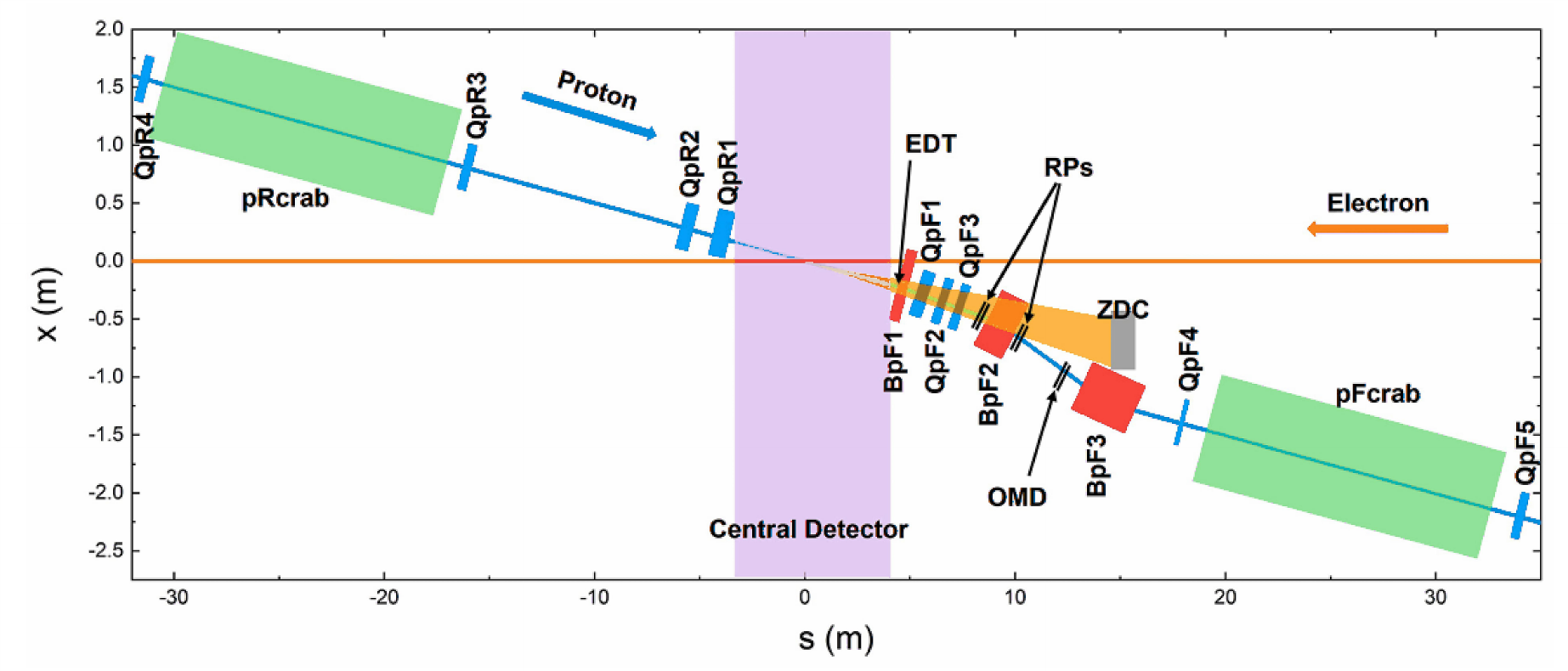}
    \caption{Schematic layout of the EicC interaction-region beam-line optics and forward detector system on the ion-beam side. The forward detector region includes the EDT, RPs, OMD, and ZDC, interleaved with dipole and quadrupole magnets.}
    \label{fig:ip_design}
\end{figure}

The realization of these forward measurements is strongly constrained by the EicC interaction-region beam-line optics and magnet layout. As illustrated in Fig.~\ref{fig:ip_design}, the ion-side forward detector package is interleaved with accelerator elements required for beam focusing and transport. Starting downstream of the central detector, this region includes the Endcap Dipole Tracker (EDT), Roman Pot (RP) stations, an Off-Momentum Detector (OMD), and finally the Zero-Degree Calorimeter (ZDC), together with the corresponding dipole (BpF) and quadrupole magnets (QpF). The EDT, located in the first forward dipole region, provides tracking and electromagnetic calorimetry for particles with polar angles of approximately 15--60~mrad around the ion beam direction, while the dipole field enables momentum analysis of forward charged particles. At smaller angles, roughly 5--15~mrad, the Roman Pot system extends the charged-particle acceptance close to the outgoing hadron beam. Off-momentum charged particles are measured by the OMD, whereas neutral particles such as photons and neutrons continue along the zero-degree direction and are detected by the ZDC.

Among the forward detector components, the Zero-Degree Calorimeter (ZDC) plays a central role in measuring neutral particles traveling along the outgoing ion-beam direction within its defined angular acceptance ($0-15$~mrad)~\cite{Anderle:2021wcy,Milton:2025zdc}. The ZDC is designed to capture very forward photons and neutrons that evade the acceptance of the main central detector. Its design is subject to stringent operational constraints: it must fit within the highly restricted spatial envelope near the outgoing hadron beam line, withstand a high-radiation environment, and deliver excellent energy and spatial resolutions. Crucially, to distinguish between different forward physics channels, such as separating high-energy photons (\textit{e.g.} from $\pi^{0}$ decay) from spectator neutrons in meson-structure studies, the ZDC must possess robust particle-identification capabilities.

The ZDC consists of two parts, an upstream electromagnetic calorimeter, followed by a hadronic calorimeter, which relies on highly segmented sampling calorimeter technologies. In this study, we evaluate a design utilizing a Spaghetti-type calorimeter (SPACAL), which embeds scintillating fibers within a dense absorber matrix~\cite{Acosta:1990qs,Armstrong:1998qs,Akchurin:2005an,Lee:2017xss,Simeonov:2022yqy}. This technology is particularly well-suited for the EicC energy regime—where typical incident neutrons possess energies around 10~GeV—due to its fine granularity of sampling and fast timing characteristics. 

In this paper, we present the geometric design and simulated physics performance of the EicC ZDC. The paper is arranged as follows: Section~\ref{sec:hadronic_calorimeter} details the software framework and the proposed detector layouts. Section~\ref{sec:simulation_result} evaluates the expected detector performance, with a specific focus on energy resolution, timing resolution, and photon/neutron PID accuracy. Section~\ref{sec:summary_and_outlook} summarizes the main findings and discusses the outlook for further optimization and detector prototyping.

\section{Hadronic Calorimeter Design}
\label{sec:hadronic_calorimeter}

\begin{figure}[htbp]
  \centering
  \includegraphics[width=1.0\textwidth]{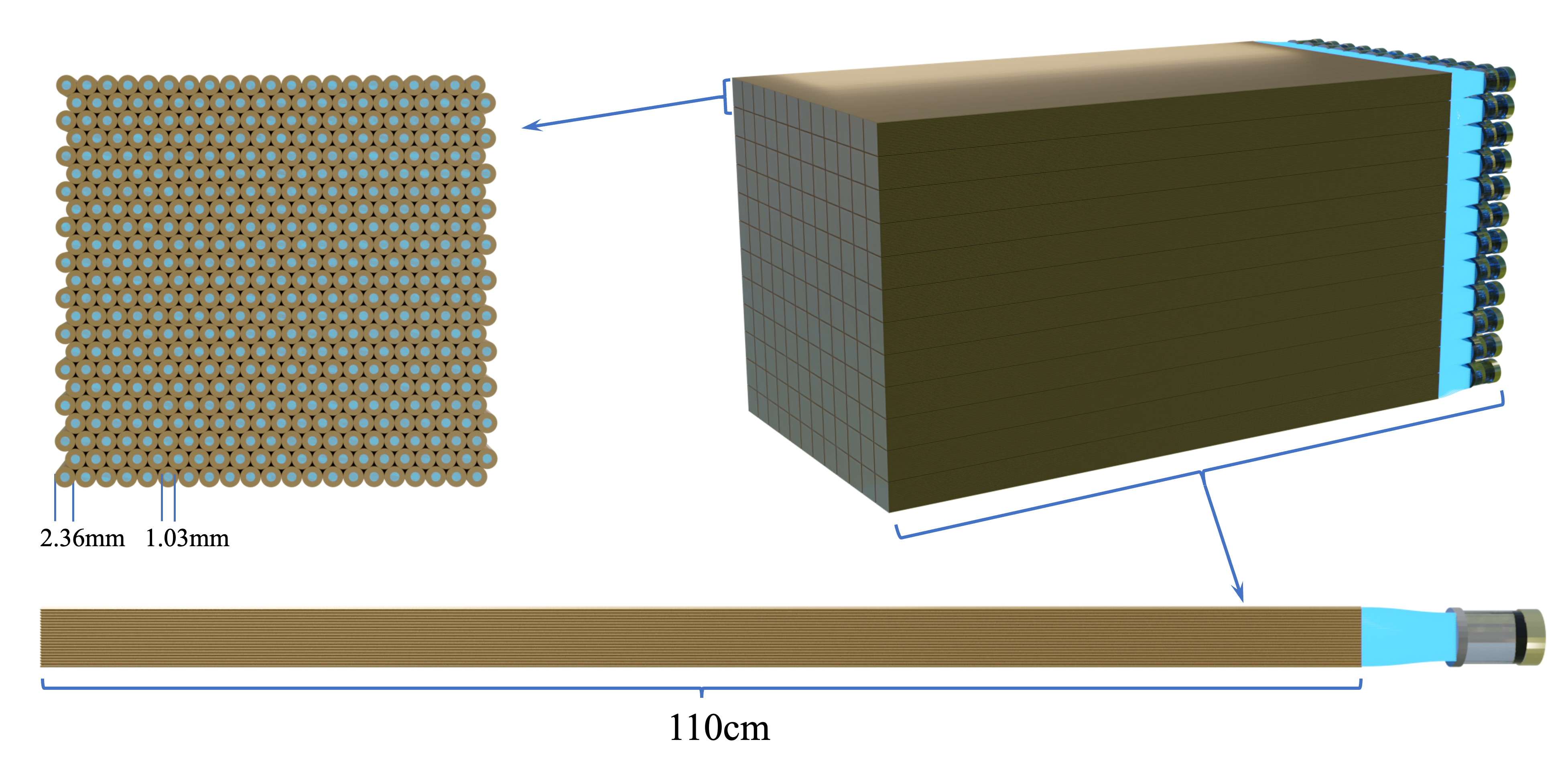}
  \caption{Schematic view of the ZDC design. The interstitial gaps between the yellow copper tubes (shown in black) are filled with a tungsten-powder/epoxy mixture.}
  \label{fig:zdc_design_view}
\end{figure}

The Spaghetti-type calorimeter (SPACAL) is adopted as the baseline design for the ZDC hadronic section due to its excellent performance in terms of compactness, radiation hardness, fast signal response, and resolution~\cite{Acosta1992LateralSP,Armstrong:1998qs,ADINOLFI2002326}. Structurally, the detector consists of a high-density absorber matrix—typically made of copper or lead—within which, a large number of thin scintillating fibers are uniformly embedded as shown in Fig.~\ref{fig:zdc_design_view}. These fibers are arranged parallel to the incident particle direction, forming a quasi-continuous active medium interleaved with passive absorber material.

When high-energy hadrons enter the calorimeter, they undergo successive elastic and inelastic interactions within the absorber, initiating hadronic showers composed of secondary particles such as pions, protons, neutrons, and electromagnetic subshowers from neutral pion decay~\cite{friend1976measurements,HOLDER197869,Cheshire1975MeasurementsOT,PhysRevD.12.2587,GRANT1975167,Fabjan:2003aq,Wigmans:2018fua}. The absorber material plays a crucial role in driving the shower development, ensuring sufficient interaction probability and containment of showers within a reasonable detector size — not only longitudinally within a limited detector length, but also transversely~\cite{WIGMANS1987389,Speckmayer:1443830}. Compared to traditional sampling calorimeters with discrete layers, the SPACAL configuration provides a much finer sampling structure due to the continuous distribution and high density of scintillating fibers, which significantly reduces sampling fluctuations~\cite{Armstrong:1998qs,ACOSTA1990193}.

The scintillating fibers act as the active medium, producing optical photons when traversed by charged particles in the shower. These photons are internally reflected along the fiber length and guided toward photodetectors such as Silicon Photomultipliers (SiPMs) or photomultiplier tubes (PMTs)~\cite{BRAVAR2024168766}. The fibers are aligned nearly parallel to the incident particle direction to channel scintillation light directly to the rear readout photodetectors, which reduces light attenuation variations and improves response uniformity. The small diameter and close packing of fibers allow high sampling frequency, which helps counteract sampling fluctuations and improves energy resolution~\cite{DESALVO1995122,LI2026171595,CARDINI201641}.

\subsection{Detector Geometry and Structure}

\begin{figure}[htbp]
  \centering
  \includegraphics[width=0.75\textwidth]{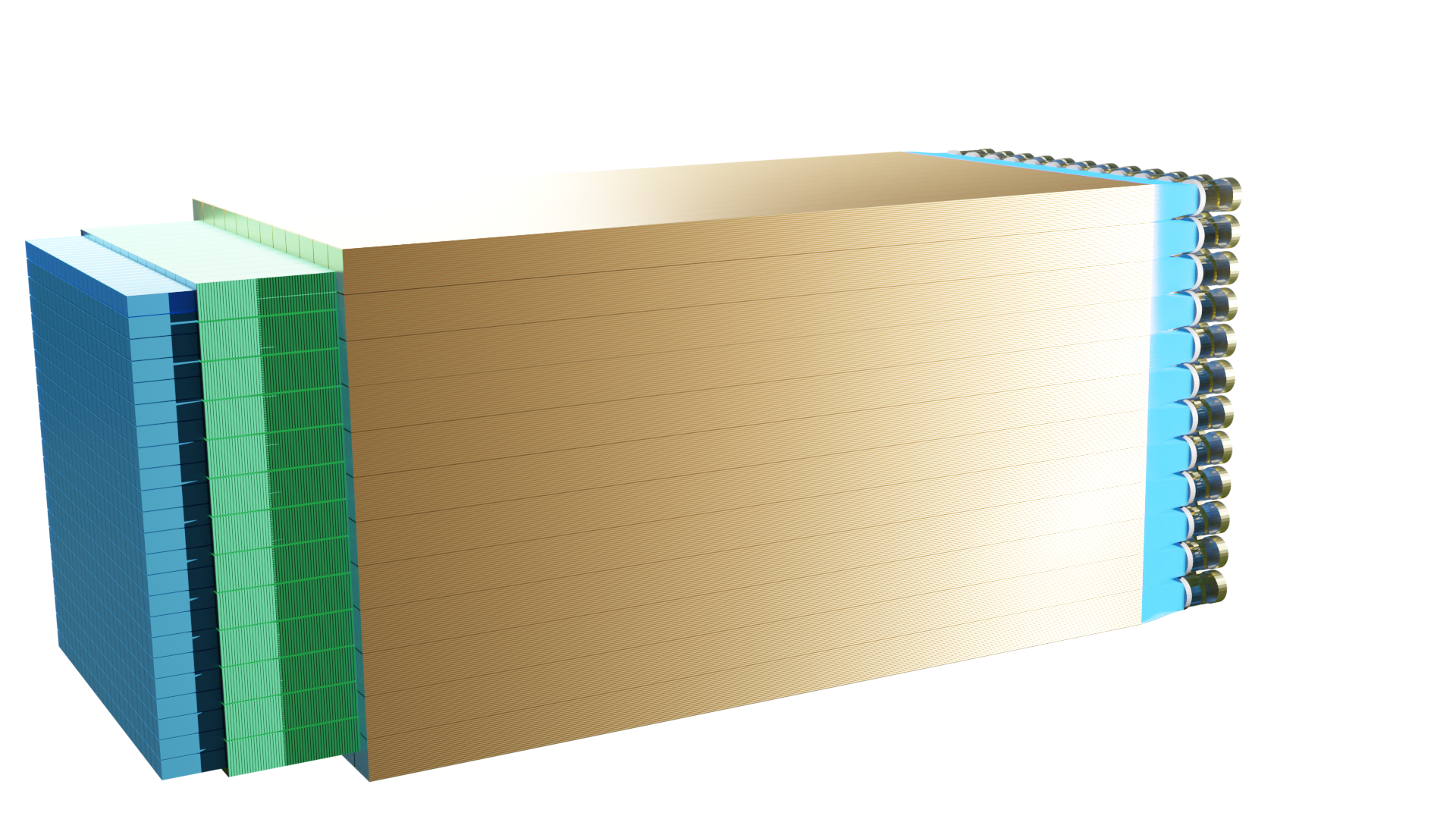}
  \caption{Schematic view of the complete ZDC calorimeter system. From left to right, the detector comprises a $23\times23$ array of PbWO$_4$ crystals, a $13\times13$ array of Shashlik modules with 50 alternating 1-mm lead and 2-mm scintillator layers, and a $12\times12$ hadronic-calorimeter array of $5\times5\times110~\mathrm{cm}^3$ modules. The photosensor assemblies are shown at the downstream end of the HCal.}
  \label{fig:emcal_design_view}
\end{figure}

The geometric design of the ZDC hadronic section is strongly constrained by the limited installation space in the EicC forward region~\cite{Anderle:2021wcy}. In the transverse direction, the detector cross section is restricted to $60 \times 60~\text{cm}^2$, as a larger size would conflict with the required beam-pipe clearance. In the longitudinal direction, the available space is approximately 2~m and is mainly limited by downstream beamline components, such as QpF4 and pFCrab, as illustrated in Fig.~\ref{fig:ip_design}. This space must accommodate both the upstream electromagnetic calorimeter (EMCal) and the downstream SPACAL HCal. As shown in Fig.~\ref{fig:emcal_design_view}, the EMCal configuration used in this study consists of an upstream PbWO$_4$ crystal calorimeter composed of a $23\times23$ array of $2.2\times2.2\times7~\mathrm{cm}^{3}$ modules, followed by a Shashlik calorimeter consisting of a $13\times13$ array of $4\times4\times15~\mathrm{cm}^{3}$ modules. Each Shashlik module contains 50 sampling layers, with each layer composed of a 1~mm thick lead absorber and a 2~mm thick plastic-scintillator layer.\footnote{The EMCal geometry adopted here is a representative configuration used to evaluate its impact on forward-neutron reconstruction and should not be interpreted as the finalized EicC EMCal design.} This leaves roughly 1.5~m for the hadronic SPACAL section, including the readout system, which is currently assumed to be based on PMTs. For a SPACAL-type calorimeter, additional fiber length is required at the rear end to reduce light loss during fiber bundling and coupling to the readout system. Therefore, the effective absorber length should be kept to about 1~m. Based on these considerations, a baseline longitudinal depth of 110~cm is adopted for the hadronic SPACAL.

\begin{figure}[htbp]
  \centering
  \includegraphics[width=0.6\textwidth]{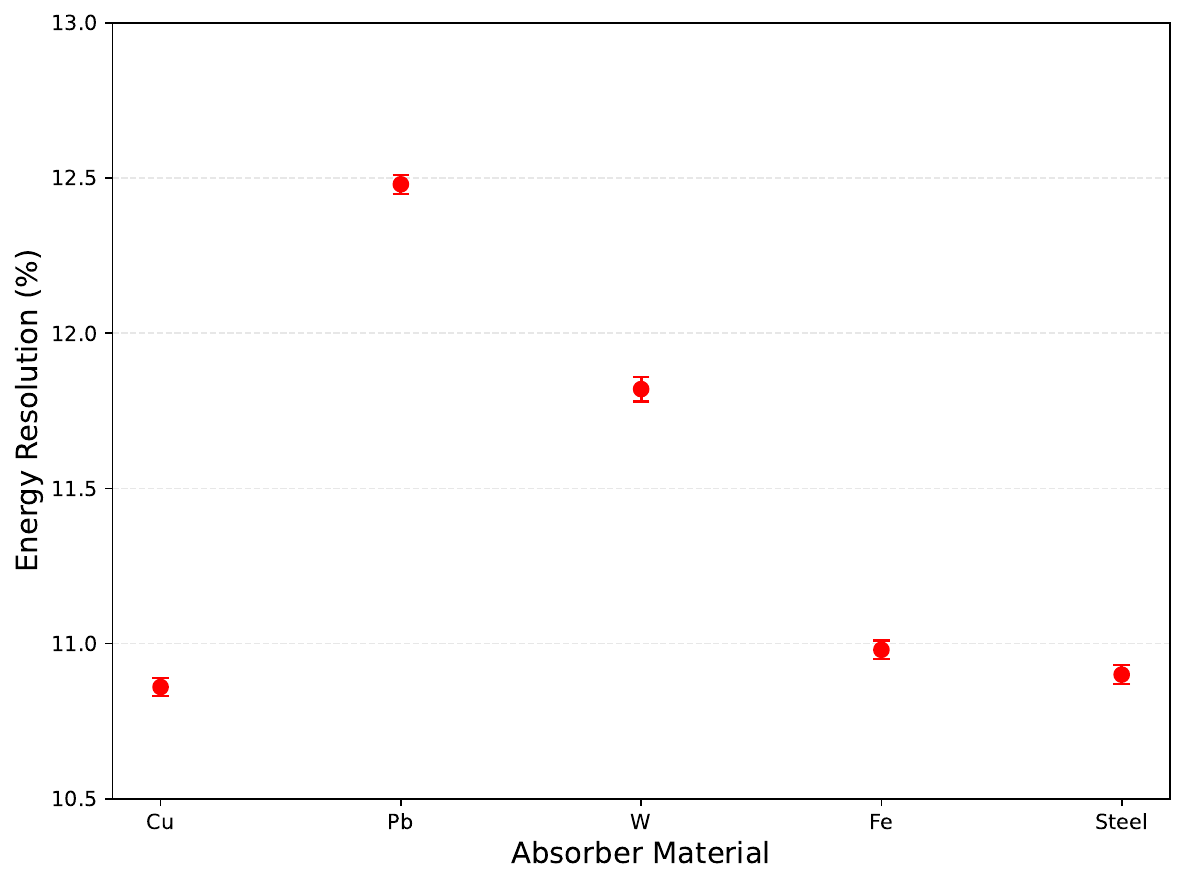}
  \caption{Comparison of the simulated energy resolution for 10 GeV incident neutrons using different absorber materials.}
  \label{fig:absorber_materials}
\end{figure}

To determine the optimal absorber material, a comparative study was carried out via the Geant4 simulation toolkit~\cite{GEANT4:2002zbu}, with candidate materials covering copper, iron, stainless steel, lead, and tungsten. As illustrated in Fig.~\ref{fig:absorber_materials}, copper, iron and stainless steel exhibit superior energy resolution relative to the remaining materials under identical geometric and optical density conditions. Nevertheless, iron possesses intrinsic magnetism, while stainless steel features high hardness and poor machinability~\cite{jmmp8060238}; both materials are technically infeasible for fabricating the SPACAL prototype. Benefiting from its outstanding energy resolution performance, alongside favorable mechanical machinability, thermal stability and wide industrial accessibility~\cite{Wigmans:1256562,Cascella_2016}, copper is therefore chosen as the baseline absorber material for the SPACAL matrix.

\begin{figure}[htbp]
  \centering
  \includegraphics[width=0.75\textwidth]{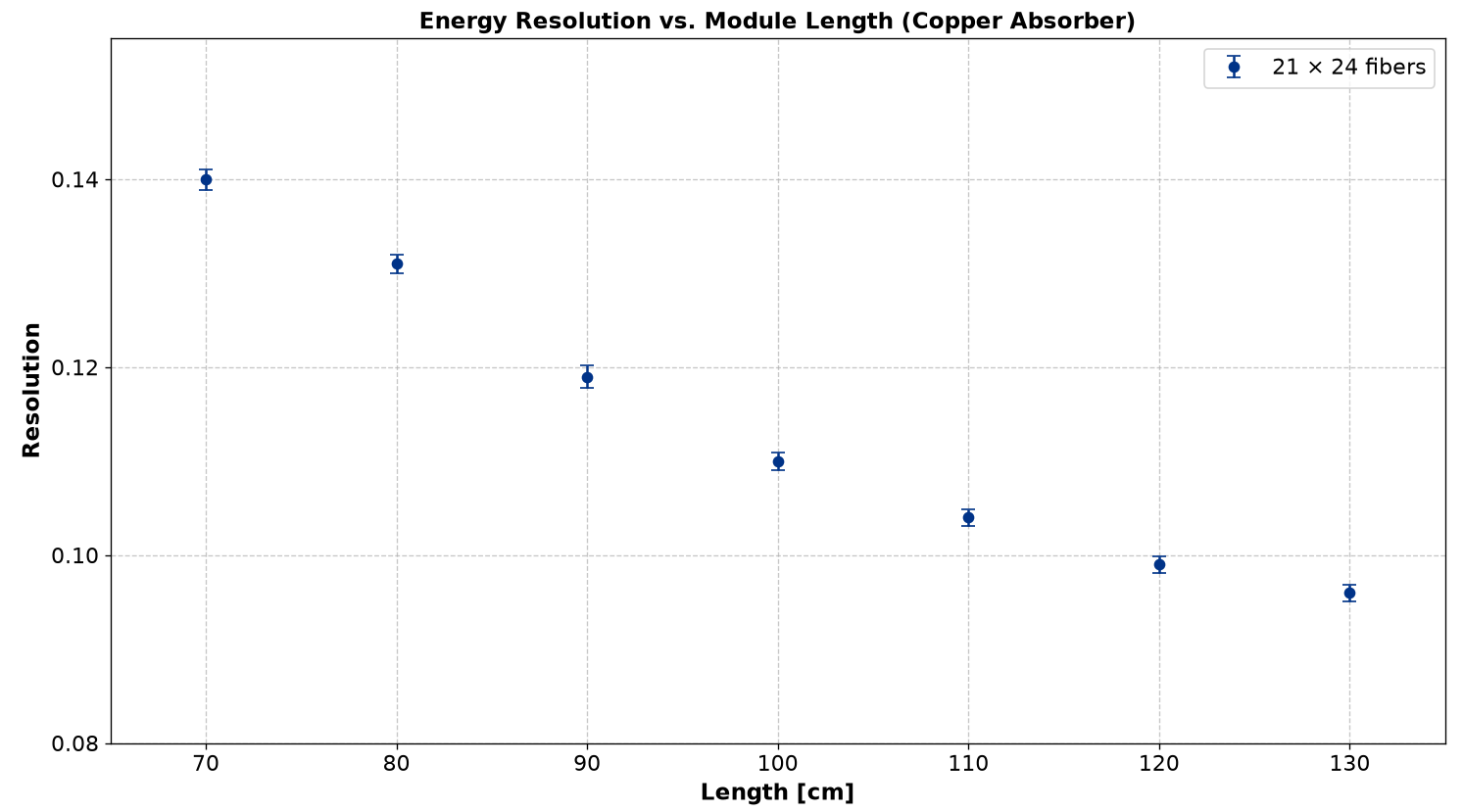}
  \caption{Simulated energy resolution as a function of module length for 10 GeV incident neutrons, with copper used as the absorber material.}
  \label{fig:reso_nfiber_length_1}
\end{figure}

\begin{figure}[htbp]
  \centering
  \includegraphics[width=0.75\textwidth]{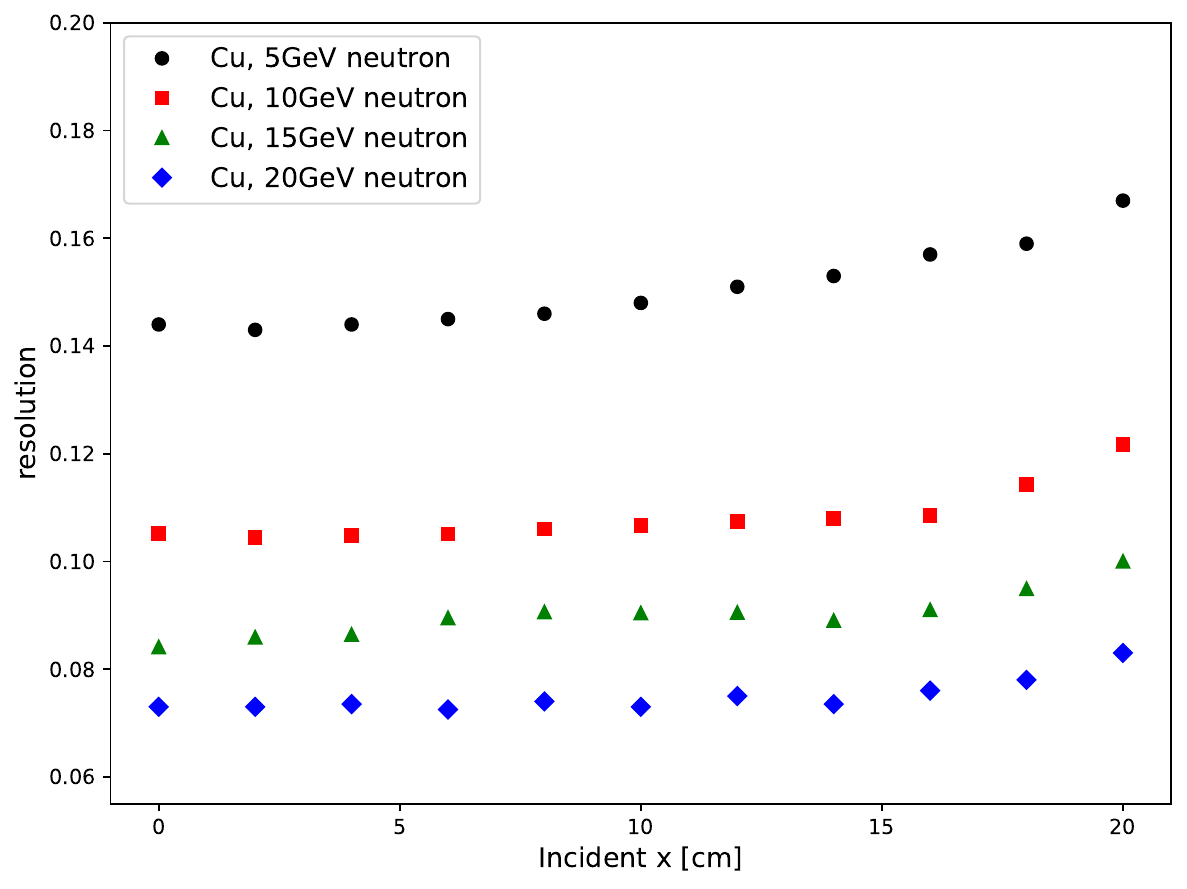}
  \caption{Spatial dependence of the simulated detector energy resolution on the neutron impact position.}
  \label{fig:copper_res_scatter_5_20}
\end{figure}

With copper selected as the absorber, systematic parameter scans were performed to validate the geometric design. The module adopts a geometric arrangement of 21×24 copper tubes. Each copper tube has an outer diameter of 2.36 mm and an inner diameter of 1.03 mm, forming an approximately square cross-section of $5 \times 5~\text{cm}^2$. The main geometrical and material parameters of the baseline HCal module are summarized in Table~\ref{tab:hcal_module}. 

\begin{table}[htbp]
    \centering
    \caption{Baseline geometry and material parameters of a single HCal SPACAL module.}
    \label{tab:hcal_module}
    \begin{tabular}{ll}
        \hline
        \textbf{Parameter} & \textbf{Specification} \\
        \hline
        Module transverse size        & $\sim 5 \times 5~\mathrm{cm}^{2}$ \\
        Module length                 & $110~\mathrm{cm}$ \\
        Absorber material             & Copper \\
        Tube arrangement              & $21 \times 24$ \\
        Number of tubes/fibers        & 504 \\
        Copper-tube outer diameter    & $2.36~\mathrm{mm}$ \\
        Copper-tube inner diameter    & $1.03~\mathrm{mm}$ \\
        Active material               & Plastic scintillating fiber \\
        Fiber diameter                & $\sim 1.0~\mathrm{mm}$ \\
        Interstitial filling material & Tungsten-powder/epoxy mixture \\
        Photodetector                 & CR285/R11102 Hamamatsu PMT \\
        \hline
    \end{tabular}
\end{table}

Because the circular tubes do not fill space completely, the interstitial gaps are filled with a tungsten-powder/epoxy mixture, which raises the effective absorber density and improves longitudinal shower containment within the compact envelope. The longitudinal depth scan, shown in Fig.~\ref{fig:reso_nfiber_length_1}, demonstrates that extending the module length beyond 110~cm yields only marginal improvements in energy resolution, consistent with previous SPACAL studies~\cite{Acosta:1990qs,Armstrong:1998qs}. The transverse acceptance was evaluated via position-dependent scans, as illustrated in Fig.~\ref{fig:copper_res_scatter_5_20}. Only the central ±20 cm region of the ZDC is considered for effective acceptance calculation, which corresponds to an angular coverage of 15 mrad. For incident neutrons from 5~GeV to 20~GeV, the energy resolution remains stable within the central fiducial region (radial displacements up to 10~cm from center). As the incident position approaches the lateral boundaries (e.g., at $x = 20$~cm), transverse shower leakage leads to modest resolution degradation—from approximately 14.5\% to 16.8\% for 5~GeV neutrons, and from 7.3\% to 8.3\% for 20~GeV neutrons~\cite{Acosta1992LateralSP}. These scans quantitatively confirm that the selected $60 \times 60 \text{ cm}^2$ design adequately contains lateral shower development while maximizing the active fiducial area. Furthermore, there are feasible approaches to further optimize the detector performance and suppress transverse energy leakage. Except for the side adjacent to the beam pipe, additional SPACAL modules can be arranged on the other lateral sides to effectively constrain shower leakage, which can significantly reduce the position dependence of the energy resolution.

\begin{figure}[htbp]
  \centering
  \includegraphics[width=1.0\textwidth]{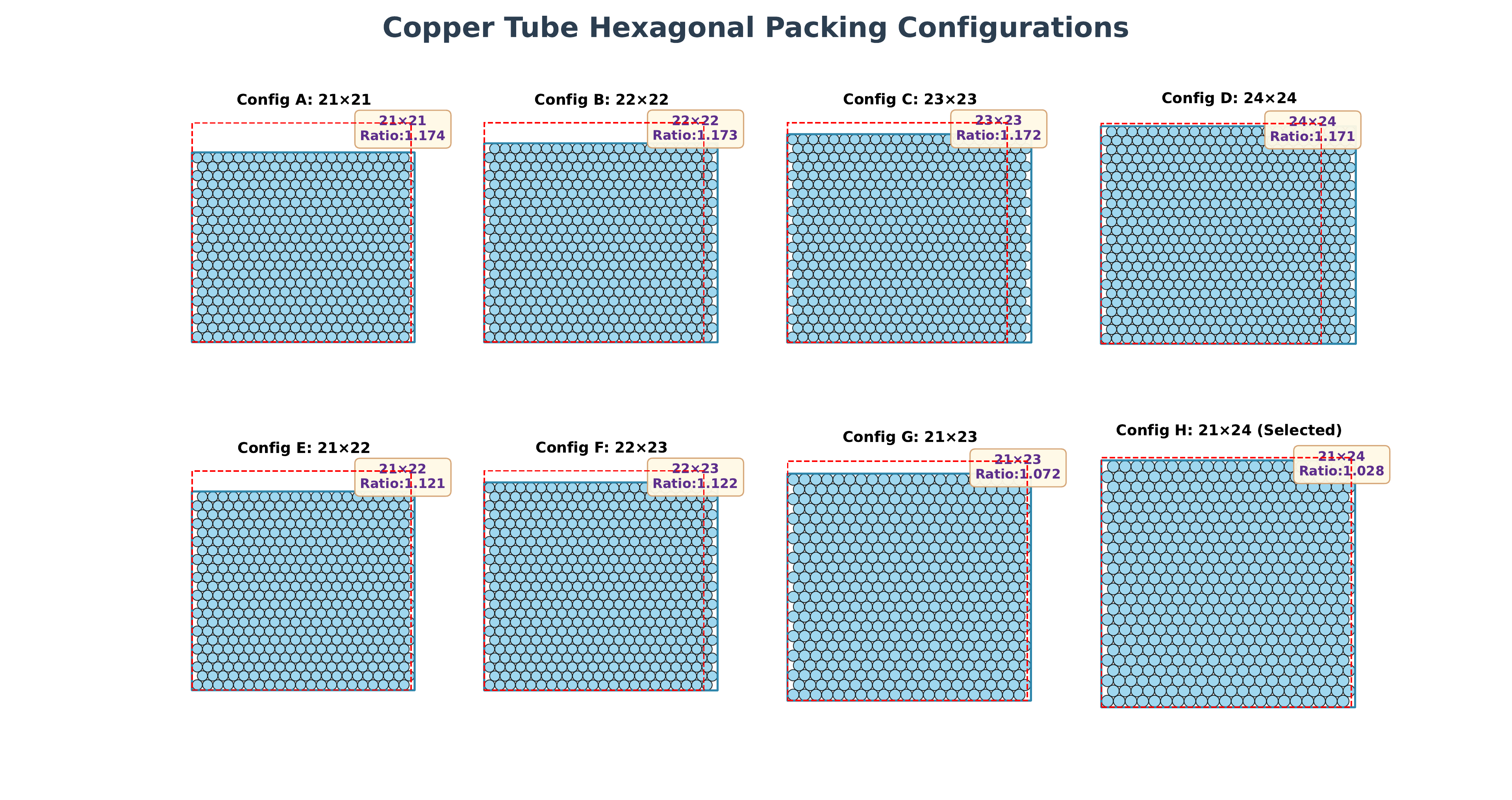}
  \caption{Comparison of different fiber array configurations in terms of geometric aspect ratio for module structural optimization. The $21\times24$ layout gives an aspect ratio close to unity while keeping the fiber count moderate, balancing near-square geometry against efficient coupling to the PMT. The red dashed-line boxes indicate the area of 5~cm $\times$ 5~cm.}
  \label{fig:tube_packing_comparison}
\end{figure}

\begin{figure}[htbp]
  \centering
  \includegraphics[width=0.70\textwidth]{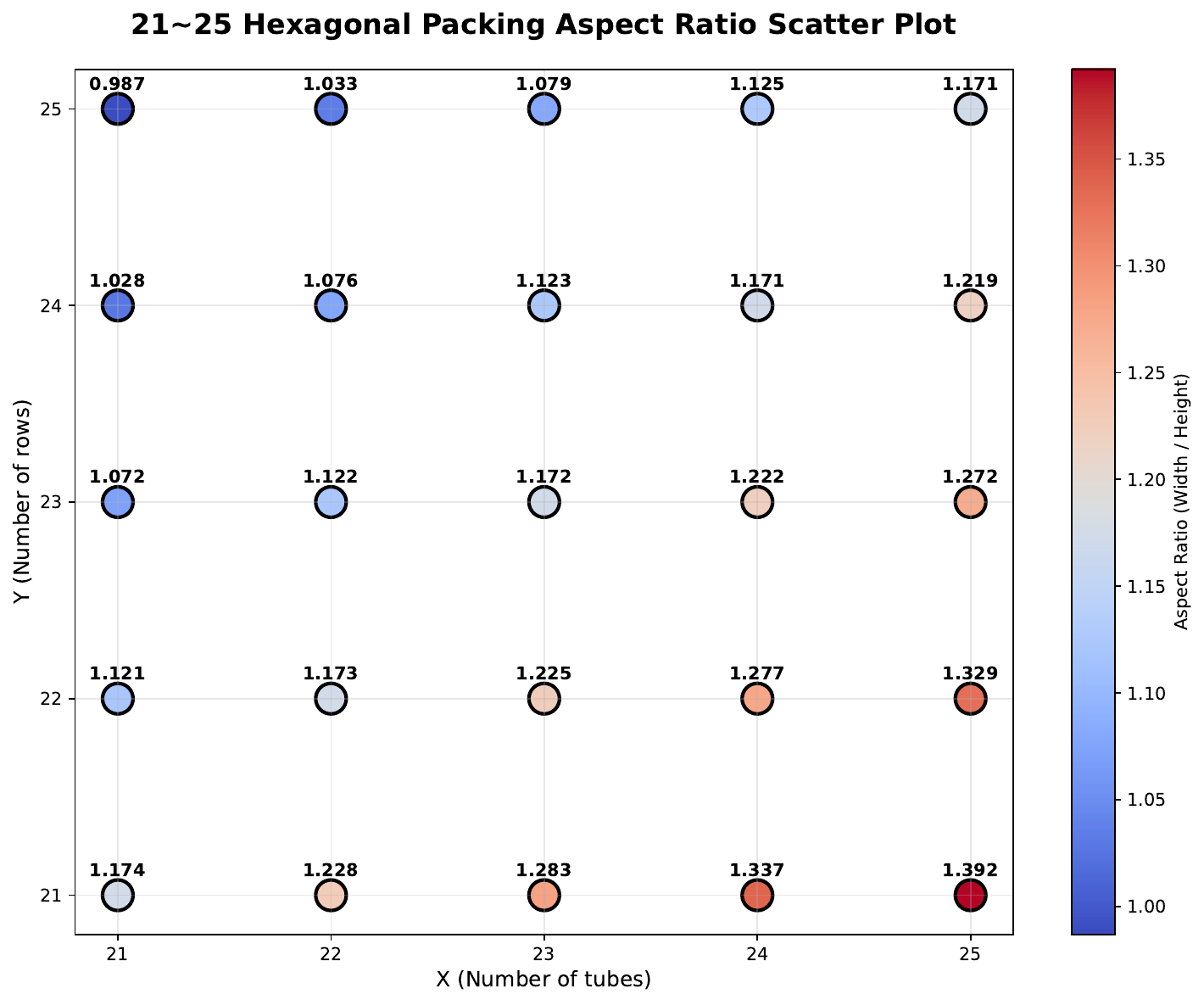}
  \caption{Width-to-height aspect ratios of the bounding boxes for hexagonally close-packed copper-tube configurations with 21--25 rows and columns. Each point represents a distinct configuration, with color indicating the corresponding aspect ratio. The $21 \times 24$ configuration achieves a near-unity aspect ratio of 1.028 while maintaining a moderate fiber count.}
  \label{fig:ratio_scatter}
\end{figure}

The resulting baseline detector is segmented into a $12 \times 12$ array of independent modules, each measuring $5 \times 5 \text{ cm}^2$. Considering the structural uniformity, especially the aspect ratio uniformity of the module, the staggered (hexagonal close-packed) arrangement of the circular tubes inherently fails to achieve a perfect 1:1 geometric ratio. To address this issue, multiple fiber configuration schemes spanning a $21\times21$ to $25\times25$ tube grid were systematically compared, as illustrated in Fig.~\ref{fig:tube_packing_comparison}. The total fiber count per module is also deliberately constrained: since the fiber bundle is directly coupled to the PMT photocathode, an excessive number of fibers would push the peripheral fibers toward the edge of the effective photosensitive area, where the photon detection efficiency is reduced. Within this constraint, the asymmetric $21 \times 24$ array is adopted, as illustrated in Fig.~\ref{fig:ratio_scatter}: its aspect ratio of 1.028 is close to unity while the fiber count stays moderate at 504. Marginally more square layouts such as $21 \times 25$ exist, but their larger fiber count increases the risk of response non-uniformity at the PMT coupling. The final dimensions and material choices, validated through systematic parameter scans, demonstrate that the proposed geometry satisfies the stringent spatial and performance requirements of the EicC ZDC as shown in Fig.~\ref{fig:zdc_design_view}.

\subsection{Optical Photon Simulation}
\label{sec:optical_design}

The optical design of the SPACAL calorimeter is critical to maximizing light collection efficiency and meeting the system’s target energy and timing resolutions. With total internal reflection, this detector guides scintillation photons directly through active plastic fibers to rear-mounted photodetectors.
In contrast to conventional designs that rely on wavelength-shifters (WLS) for intermediate photon transmission, this direct optical path removes the secondary photon re-emission process in WLS, substantially reduces optical photon loss, and greatly shortens signal formation time. In the high-rate experimental environment of EicC, the excellent ultra-fast timing response enables efficient discrimination of various forward collision events and brings remarkable performance benefits.

\begin{figure}[htbp]
    \centering
    \begin{minipage}[t]{0.48\textwidth}
        \centering
        \includegraphics[width=\linewidth]{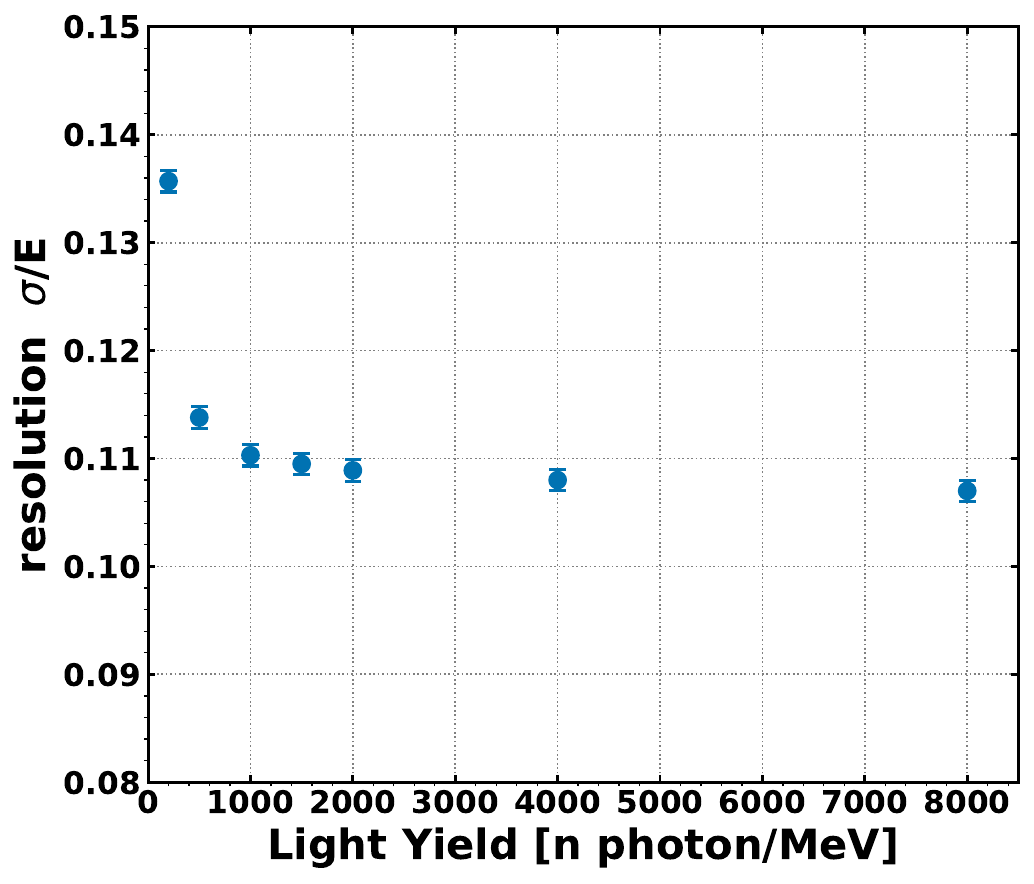}
        \caption{Simulated energy resolution as a function of scintillating-fiber light yield for 10 GeV incident neutrons, assuming a fiber attenuation length of 3.28 m.}
        \label{fig:spahgetti_ly}
    \end{minipage}
    \hfill
    \begin{minipage}[t]{0.48\textwidth}
        \centering
        \includegraphics[width=\linewidth]{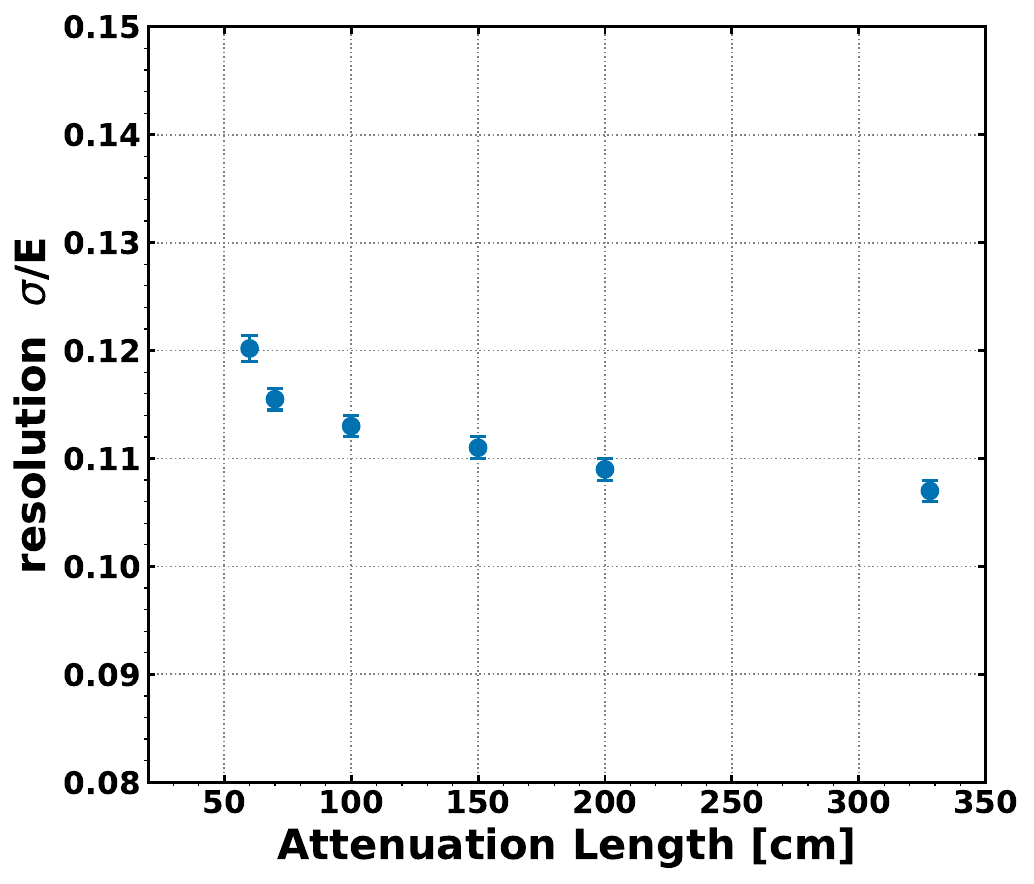}
        \caption{Simulated energy resolution as a function of fiber attenuation length for 10 GeV incident neutrons, assuming a scintillating-fiber light yield of 8,000 photons/MeV.}
        \label{fig:spahgetti_al}
    \end{minipage}
\end{figure}

To quantitatively optimize the fiber specifications, the impacts of intrinsic light yield and attenuation length were evaluated via Geant4 simulations.
As shown in Fig.~\ref{fig:spahgetti_ly}, the reconstructed energy resolution improves rapidly at low light yields but plateaus, becoming relatively insensitive to intrinsic light yield once it exceeds a light yield of approximately 2000 photons/MeV. Because standard commercial scintillating fibers (e.g., Kuraray SCSF-78 or SCSF-81) routinely achieve yields of 7000--8000 photons/MeV, intrinsic light yield does not limit the performance of this design~\cite{Bravar:2022xbz,Losekamm:2024yvw}.
Conversely, the attenuation length is a much more critical parameter due to the 110 cm longitudinal depth of the modules. As illustrated in Fig.~\ref{fig:spahgetti_al}, energy resolution exhibits a steep, continuous improvement as the attenuation length increases from 50 cm to roughly 350 cm, after which the performance gains begin to diminish. As a result, selecting high quality commercial fibers with attenuation length approaching or exceeding 4 m would be desirable in order to achieve high energy resolution. This ensures sufficient uniformity in the longitudinal light collection to mitigate shower depth dependent signal fluctuations.

The optical readout scheme is also tightly constrained by the physical geometry of the EicC interaction region. Legacy SPACAL designs, such as the BNL-E864 hadronic calorimeter~\cite{Armstrong:1998qs}, employed $10\times10~\text{cm}^2$ modules that required $\sim42~\text{cm}$ long adiabatic light guides and bulky PMTs. This resulted in a total assembly length of nearly $1.9~\text{m}$, which would exhaust the available space in the EicC ZDC region and preclude the installation of the upstream electromagnetic calorimeter. By reducing the module cross-section to $5\times5~\text{cm}^2$ with a similar fiber density, the fiber bundle can be directly coupled to compact, large-area photodetectors (such as PMT arrays). This direct-coupling configuration eliminates the need for spatial light guides, drastically reducing the total detector length while preserving a high fiber density and photon statistics established in the geometric design.

\begin{figure}[htbp]
  \centering
  \includegraphics[width=0.8\textwidth]{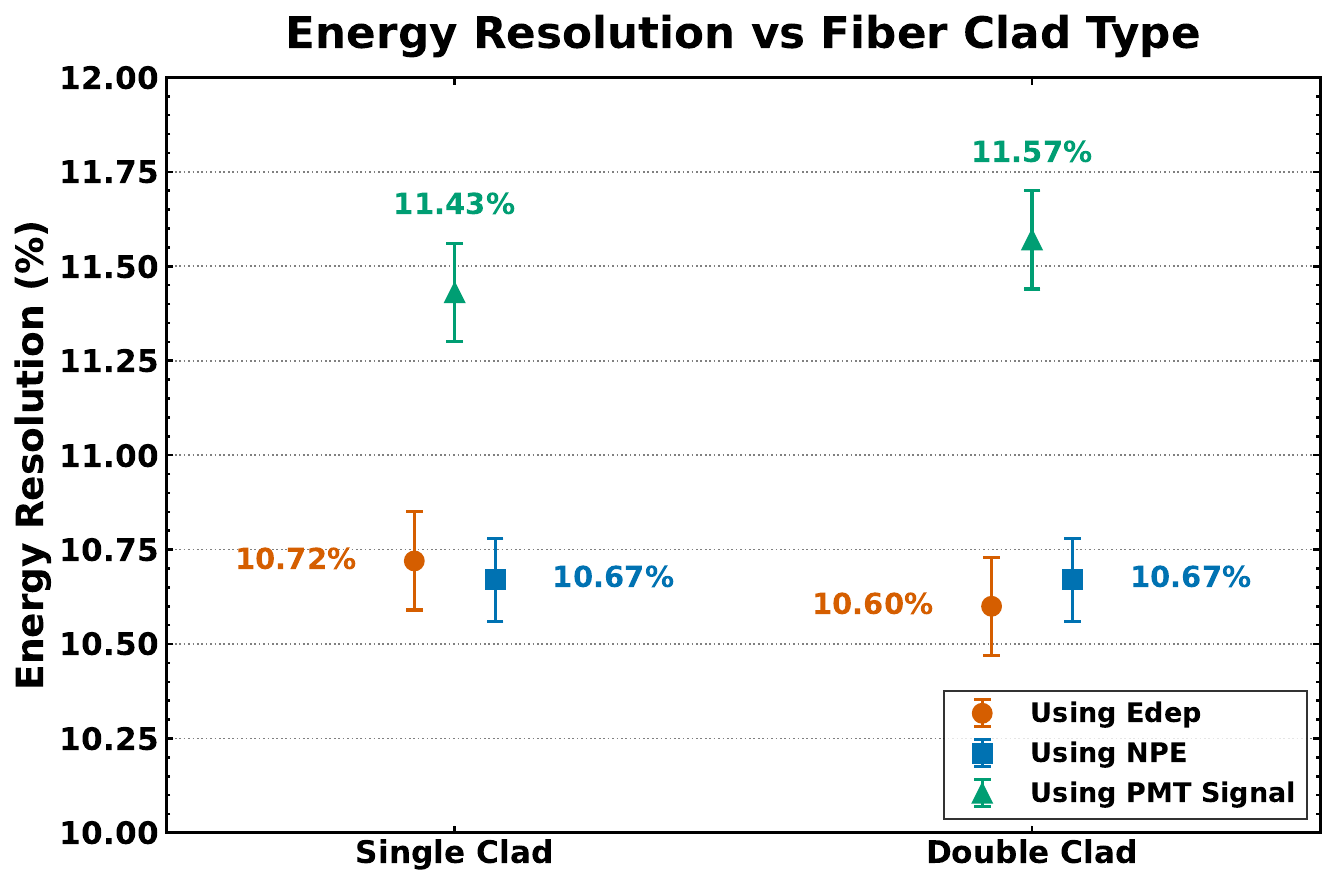}
   \caption{Comparison of the energy resolution of single-clad and double-clad scintillating fibers under three energy reconstruction methods, for incident 10 GeV neutrons. 
   }
  \label{fig:spahgetti_fiberclad}
\end{figure}

Finally, the impact of the fiber cladding structure on light collection efficiency was investigated. Double-clad fibers feature an additional intermediate layer that increases the photon capture efficiency at large angles. However, comprehensive simulations tracking the signal from raw energy deposition to the fully digitized PMT readout (Figure~\ref{fig:spahgetti_fiberclad}) reveal no operational advantage for the double-clad configuration. The energy resolution---whether derived from the raw deposited energy ($E_{\mathrm{dep}}$), the photoelectron yield (NPE), or the fully digitized PMT signal---differs between the two fiber types by no more than $\sim$0.2\% in absolute terms, within the statistical uncertainty of the simulation. This is expected because the design already operates above the light-yield plateau (Fig.~\ref{fig:spahgetti_ly}); therefore, the additional light collected by the double-clad structure provides no measurable improvement in energy resolution, despite the approximately twofold increase in collected light. Given that double-clad fibers typically cost $30\%\text{--}40\%$ more without providing a measurable performance enhancement in the final digitized signal, single-clad fibers provide a highly cost-effective solution. Consequently, high-attenuation, single-clad fibers directly coupled to compact photodetectors are conclusively selected as the baseline optical configuration for the prototype design.

\subsection{Signal Digitization}
\label{sec:readout_scheme}

To accurately assess the detector's performance under realistic experimental conditions, a simple digitization framework was developed to model the complete PMT readout chain. The goal of this simulation is to bridge the gap between idealized Geant4 energy deposits and experimentally measurable signals, ensuring consistency between the simulated response and true detector behavior. The framework executes a sequential transformation: from raw energy deposition ($E_{\text{dep}}$) to the number of detected photoelectrons (NPE), followed by PMT single-photon pulse superposition, analog waveform generation with realistic noise, and concluding with signal integration for the final reconstructed energy.

The conversion from simulated NPE to an analog electrical signal is governed by a PMT response model driven by empirical laboratory measurements. Initially, individual photon arrival times are extracted from Geant4 and mapped onto a time axis. These discrete photoelectron timestamps are then convolved with a measured single-photoelectron (SPE) pulse shape. To maximize the fidelity of the simulation, experimental noise profiles are superimposed onto simulated readout signals. These noise sources cover single-photon peak amplitude fluctuations, timing jitter and baseline electronics noise. Incorporating these effects enables the resultant waveforms to reproduce statistical signal fluctuations and practically shaped electronic pulses.

\begin{figure}[htbp]
    \centering
    \begin{minipage}[t]{0.48\textwidth}
        \centering
        \includegraphics[width=\linewidth]{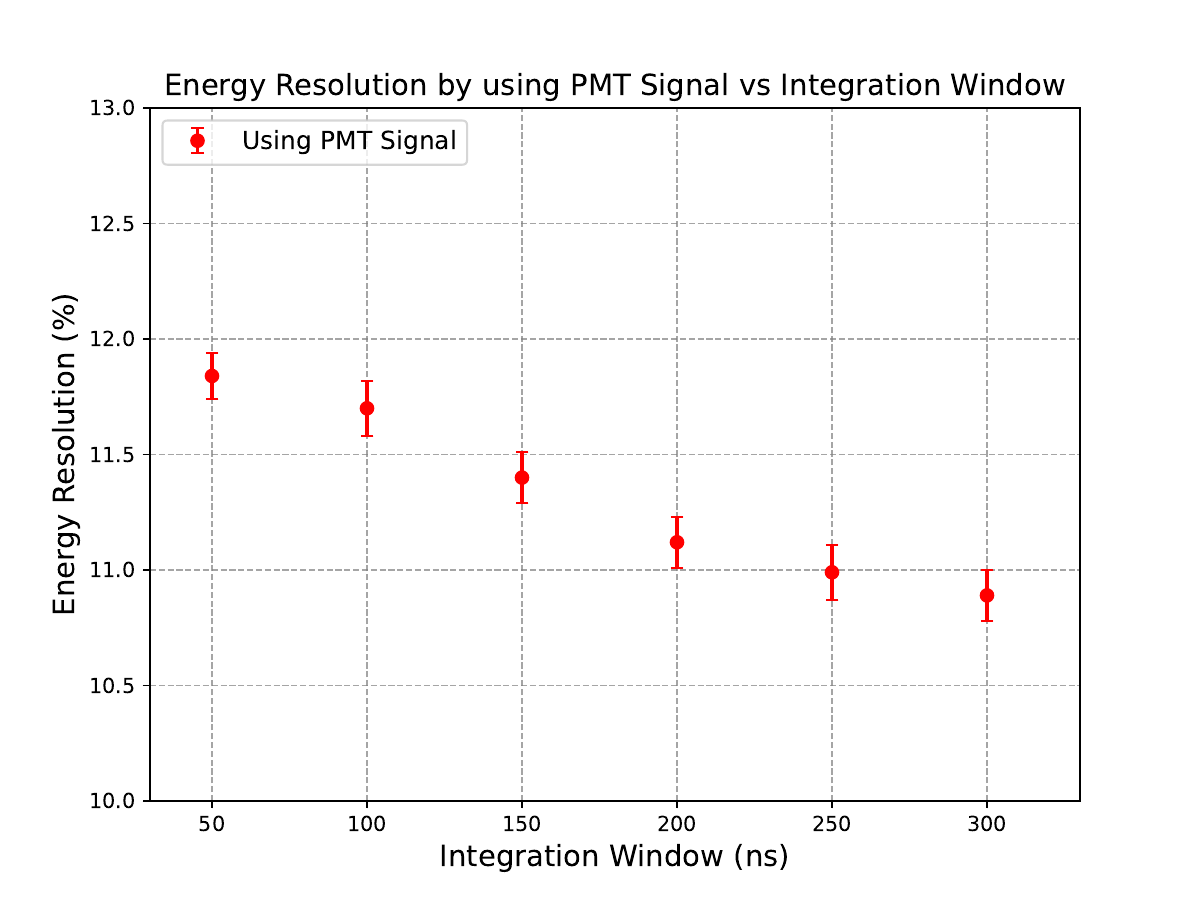}
        \caption{The effect of different integration windows on the reconstructed energy resolution of PMT signals induced by 10 GeV neutrons.}
        \label{fig:time_window}
    \end{minipage}
    \hfill
    \begin{minipage}[t]{0.48\textwidth}
        \centering
        \includegraphics[width=\linewidth]{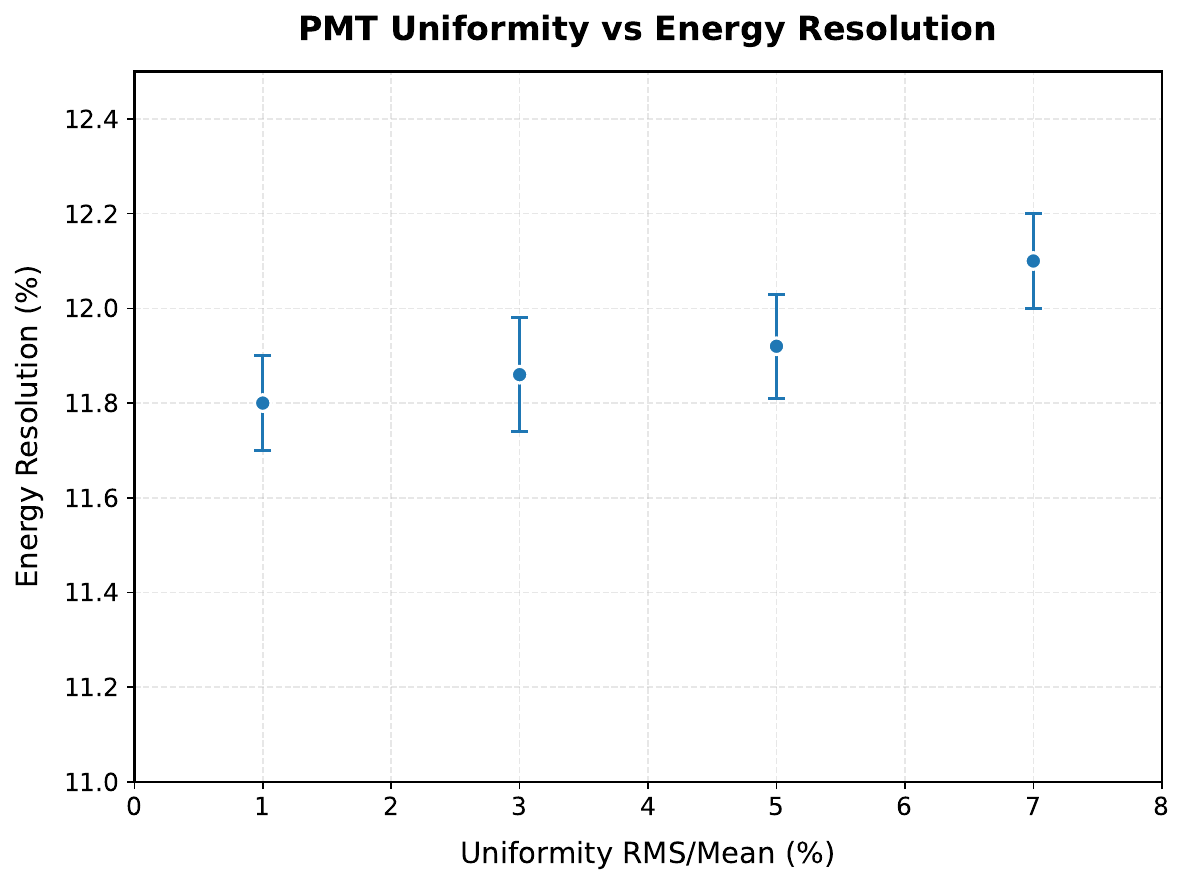}
        \caption{Correlation between Hamamatsu CR285 PMT spatial uniformity and energy resolution for 10 GeV incident neutrons.}
        \label{fig:PMT_uniformity_vs_resolution}
    \end{minipage}
\end{figure}

Following waveform generation, signals are processed via a simulated front-end electronics chain. The total signal charge is extracted by integrating the waveform within a predefined time window, adopting the standard methodology employed in high-rate data acquisition (DAQ) systems. The choice of integration window directly affects the accuracy of energy reconstruction. To optimize the energy resolution, we systematically evaluated integration windows with different durations, as illustrated in Fig.~\ref{fig:time_window}.
The results show that the reconstructed energy resolution improves gradually as the integration window is widened, since a longer window captures a larger fraction of the late-arriving scintillation light and thus more of the total signal charge. This improvement is modest, however---broadening the window from 100~ns to 300~ns lowers the resolution by only about 0.8\% in absolute terms---so an integration window as narrow as 100~ns is already sufficient to fulfill the detector design requirements. A short window of this order is also favorable for suppressing pile-up in the high-rate EicC environment.

Based on the geometrical optimization, optical performance simulations, and readout electronics modeling described above, systematic simulation studies were further carried out on an actual photoelectric conversion device. Considering the tight spatial constraints of the detector, the photon collection efficiency at the fiber ends, and the timing response characteristics, the Hamamatsu CR285 photomultiplier tube (PMT) was selected as the reference photodetector. This model features a minimum effective photocathode diameter of 34 mm for properly sized active area and a typical rise time of 2.8 ns enabling fast time response, alongside mature commercial availability, making it highly compatible with the direct-coupling optical readout scheme at the rear end of the SPACAL.

In the established digitization framework, the measured noise parameters—including a baseline noise of 0.586 mV, a single-photoelectron timing spread of 0.85 ns (FWHM), and signal amplitude fluctuations—were already incorporated. On this basis, dedicated simulations were performed to evaluate the response non-uniformity at different positions of the CR285 PMT photocathode. Photocathode response uniformity is one of the key factors affecting the energy and position resolution of the readout system. Typical uniformity requirements for PMT channels in large-scale scientific facilities are within 10\%~\cite{FUKUDA2003418,Anfimov_2017,Abusleme_2022,Ma_2022}, while high-precision experiments often demand stricter levels.

To comprehensively assess the impact of PMT photocathode response non-uniformity on the overall ZDC performance, four levels of channel-response variation—7\%, 5\%, 3\%, and a near-ideal 1\%—were simulated under otherwise identical conditions. The spatial response pattern used to model the photocathode non-uniformity was based on laboratory measurements of CR285 PMTs~\cite{ZHANG2017429,Zhang:2017LHAASO}. The results are shown in Fig.~\ref{fig:PMT_uniformity_vs_resolution}. In the near-ideal 1\% case, no noticeable impact on the resolution was observed, and even under the conservative 7\% uniformity condition, the resolution degrades only marginally from this ideal value. This indicates that at the current levels of optical and electronic noise, PMT non-uniformity at the 7\% level does not constitute a performance bottleneck, and the CR285 PMT is thus fully adequate for the required performance.

According to laboratory measurements~\cite{ZHANG2017429,Zhang:2017LHAASO}, the response uniformity of the Hamamatsu CR285 PMT is typically better than 3\%, which is well below the 10\% threshold. Therefore, this PMT solution not only satisfies the basic performance requirements of the EicC ZDC but also provides ample engineering margin.  In addition, this finding can serve as a reference for PMT selection during subsequent prototype development.

\section{Simulation Result}
\label{sec:simulation_result}

With the baseline detector geometry optimized via Geant4 simulations and the digitization framework, we now evaluate the fully realistic physics performance of the ZDC.
Two distinct simulation configurations are adopted in this section for categorized performance assessment:
(1) standalone SPACAL hadronic calorimeter (HCal) without upstream EMCal, to characterize intrinsic HCal performance including energy resolution and timing resolution;
(2) the full ZDC system integrating upstream EMCal and downstream HCal, to evaluate system-level transverse position reconstruction and particle identification (PID) utilizing shower profile information across all sectors.
In the following text, the four core metrics are grouped into two subsections corresponding to the above two configurations, providing quantitative guidance for both the engineering prototype construction and subsequent EicC physics analyses.

\subsection{Intrinsic Performance of Standalone SPACAL HCal}
\label{sec:hcal_intrinsic_performance}
This subsection evaluates the intrinsic performance of the standalone SPACAL hadronic calorimeter without upstream EMCal, including energy reconstruction resolution and timing resolution based solely on PMT waveforms from the HCal readout chain.

\subsubsection{Energy Reconstruction and Resolution}
\begin{figure}[htbp]
\centering
\begin{subfigure}[t]{0.32\textwidth}
\centering
\includegraphics[width=\linewidth]{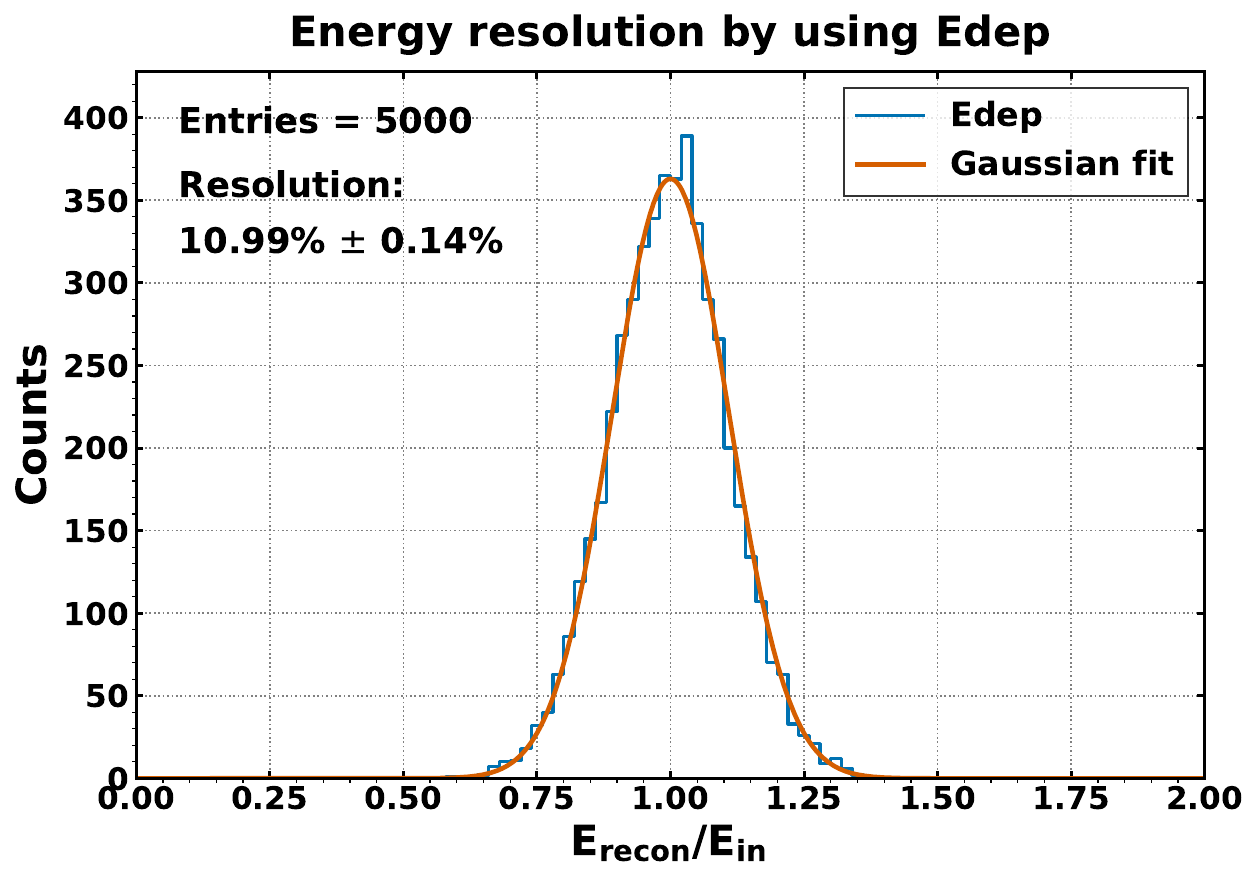}
\caption{Energy deposition ($E_{\mathrm{dep}}$)}
\end{subfigure}
\hfill
\begin{subfigure}[t]{0.32\textwidth}
\centering
\includegraphics[width=\linewidth]{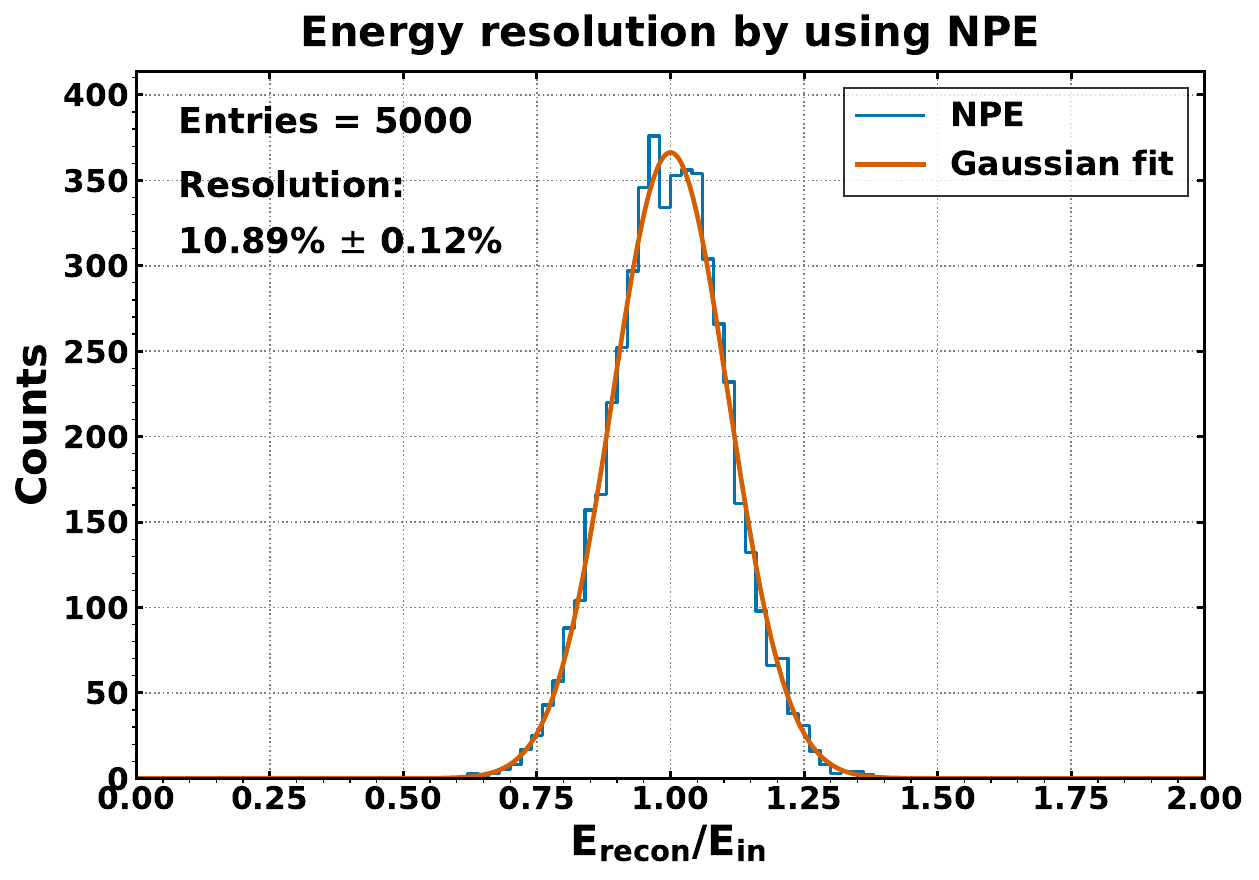}
\caption{Photoelectron yield (NPE)}
\end{subfigure}
\hfill
\begin{subfigure}[t]{0.32\textwidth}
\centering
\includegraphics[width=\linewidth]{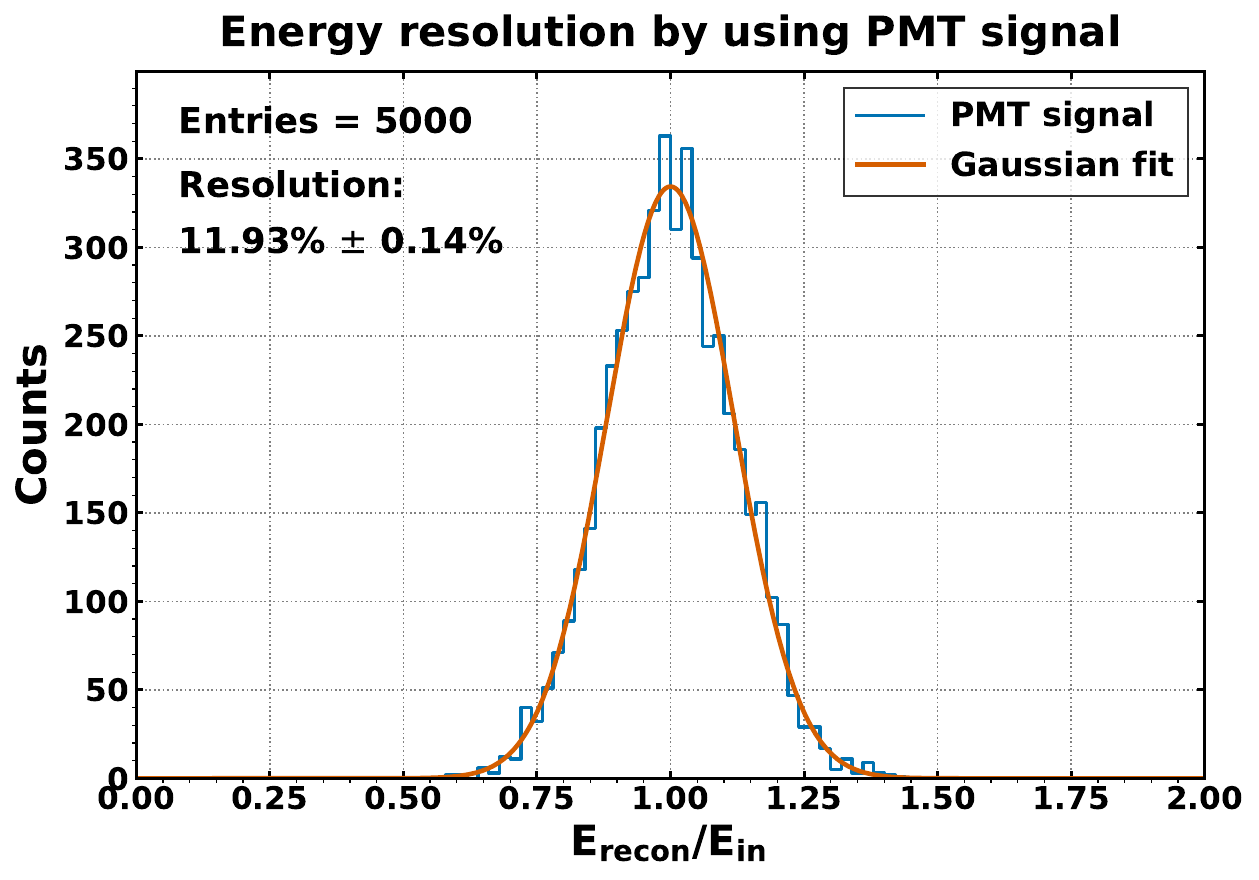}
\caption{Reconstructed PMT signal}
\end{subfigure}
\caption{The energy reconstruction chain for 10~GeV incident neutrons. The panels demonstrate the simulated detector response at sequential stages: (a) raw energy deposition in the calorimeter, (b) the detected photoelectron distribution after optical transport, and (c) the digitized PMT signal after full electronics response simulation.}
\label{fig:signal_chain_10GeV}
\end{figure}

This energy resolution study adopts a standalone HCal simulation configuration, where the upstream EMCal is excluded. Building upon the digitization framework described in Section~\ref{sec:readout_scheme}, the energy resolution was systematically evaluated by tracking the signal degradation through the multi-stage digitization chain. Fig.~\ref{fig:signal_chain_10GeV} illustrates this progression for a nominal 10~GeV incident neutron, displaying the reconstructed energies based on the raw energy deposition ($E_{\mathrm{dep}}$), the corresponding photoelectron yield (NPE), and the fully reconstructed PMT waveform.

For 10~GeV incident neutrons, the intrinsic energy resolution derived purely from the Geant4 energy deposition is $(10.99 \pm 0.14)\%$. After simulating the optical transport and converting the light yield to photoelectrons, the resolution remains highly consistent at $(10.89 \pm 0.12)\%$. This indicates that the high light yield and direct photon transport of the SPACAL design prevent photon statistics from dominating the intrinsic hadronic shower fluctuations. Finally, after passing through the complete PMT waveform generation—which superimposes realistic electronic noise, timing jitter, and signal integration—the fully reconstructed energy resolution modestly broadens to $(11.93 \pm 0.14)\%$. This $\sim1\%$ degradation mainly originates from the selection of the integration window and minor contributions induced by electronic noise.

\begin{figure}[htbp]
  \centering
  \includegraphics[width=0.85\textwidth]{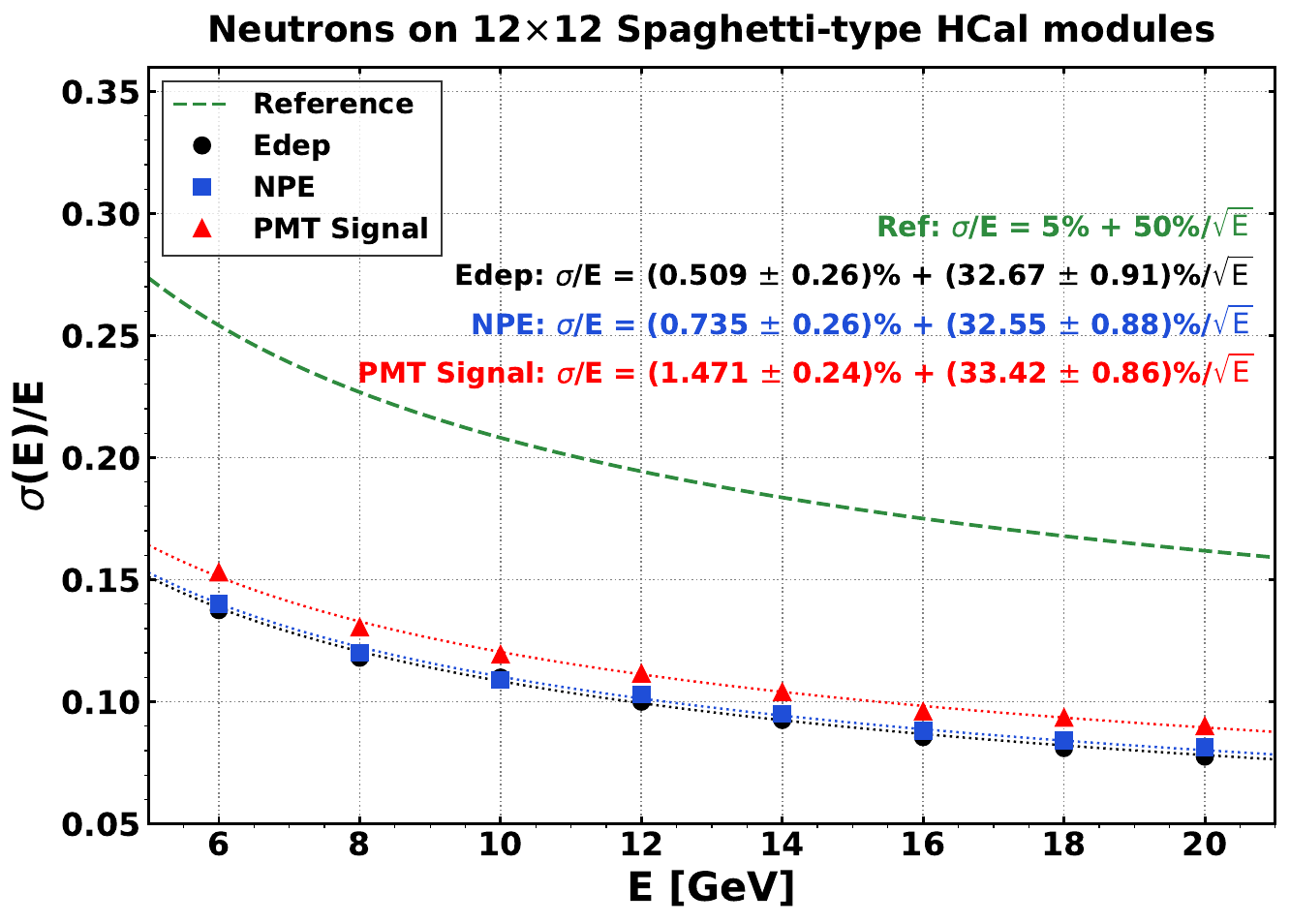}
  \caption{Energy resolution as a function of incident neutron energy. The resolutions derived from the raw energy deposition ($E_{\mathrm{dep}}$), photoelectron yield (NPE), and fully reconstructed PMT signals are compared. The solid curves represent fits to the stochastic and constant term parameterization, while the green dashed line represents the baseline performance reference.}
  \label{Tube_resolution_fit_Edep_NPE_PMT}
\end{figure}

To evaluate the detector's energy resolution across the relevant EicC kinematic regime, an energy scan spanning 6 to 20~GeV was performed. Fig.~\ref{Tube_resolution_fit_Edep_NPE_PMT} presents the energy resolution as a function of incident neutron kinetic energy, $E$, for the three sequential stages of the reconstruction chain. For many hadronic calorimeters, the energy resolution can be well described by a linear sum of a stochastic term and a constant term~\cite{Armstrong:1998qs,Leroy:2000mj,Lee:2017oye}, $\frac{\sigma_E}{E} = \frac{a}{\sqrt{E}} + b$. We therefore adopt the same parameterization to fit the data. As the signal propagates from the raw energy deposition ($E_{\mathrm{dep}}$) to the photoelectron yield (NPE), the stochastic term remains highly stable ($\sim 32.7\%$), indicating excellent optical transport. Upon applying the full electronic readout chain, the fully reconstructed PMT signal yields an extracted stochastic term of $a = (33.42 \pm 0.86)\%$ and a constant term of $b = (1.47 \pm 0.24)\%$. Furthermore, the simulated resolutions are compared against a baseline reference requirement of $50\%/\sqrt{E} + 5\%$. The proposed SPACAL design systematically outperforms this reference across the entire evaluated energy range. These results definitively demonstrate that even after the inclusion of full electronics noise, the achieved energy resolution comfortably exceeds the stringent design requirements of the EicC forward physics program.

\begin{figure}[htbp]
  \centering
  \includegraphics[width=0.85\textwidth]{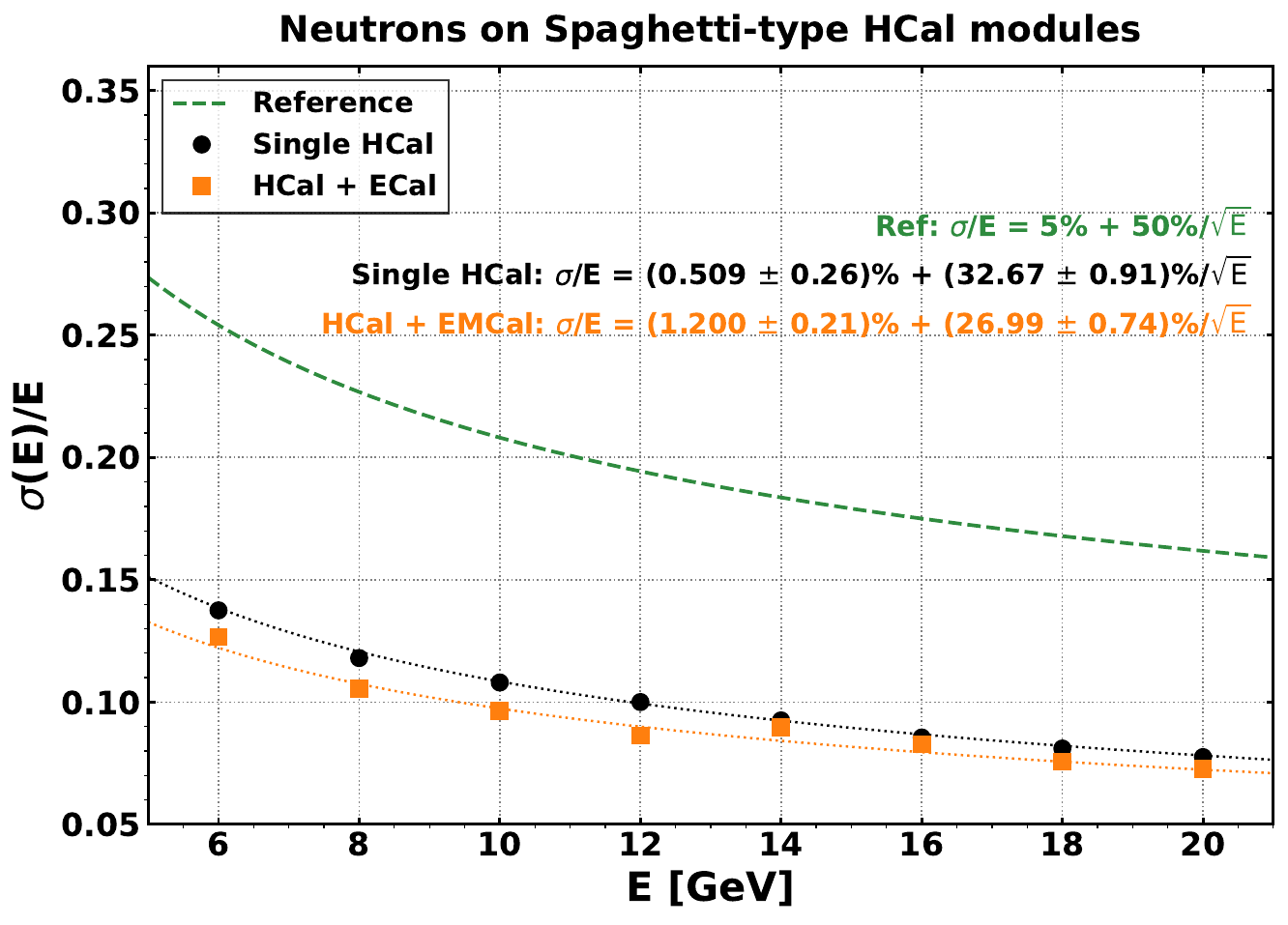}
  \caption{Fitted energy resolution curves of the standalone HCal configuration and the full coupled EMCal+HCal ZDC system.}
  \label{fig:reso_EMcal_compare}
\end{figure}

To evaluate the potential impact of the upstream EMCal on the hadronic energy resolution, an additional energy scan was performed with the full EMCal configuration included. As shown in Fig.~\ref{fig:emcal_design_view}, the EMCal is located immediately upstream of the HCal and consists of a PbWO$_4$ sector followed by a Shashlik sector. Although the energy deposited by hadrons in the EMCal is small compared with that in the HCal, it is not negligible. The reconstructed hadron energy is therefore obtained from the combined response of all three ZDC sectors:
\begin{equation}
E_{\rm{recon}} = p_{0} \cdot S_{\rm{PWO}} + p_{1} \cdot S_{\rm{Shash.}} + p_{2} \cdot S_{\rm{HCal}},
\end{equation}
where $S$ denotes the signal measured in a given ZDC sector, which may correspond to the deposited energy, NPE, or digitized waveform response. The coefficients $p_{0}$, $p_{1}$, and $p_{2}$ are free parameters determined through a linear fit. The resulting energy resolutions with and without the EMCal are compared in Fig.~\ref{fig:reso_EMcal_compare}. A slight improvement is observed when the EMCal response is included, demonstrating that the presence of the upstream EMCal does not degrade the hadronic energy reconstruction. Instead, incorporating the energy deposited in the EMCal provides a modest improvement in the overall ZDC energy resolution.

\subsubsection{Timing Resolution}
\label{sec:timing_resolution}
Timing performance is characterized using only digitized waveforms from the SPACAL HCal, with no timing information extracted from the upstream EMCal readout system.
Excellent timing resolution is an important requirement for the EicC ZDC, providing the ability to cleanly assign forward-going neutral particles to their correct bunch crossings and mitigate high-rate pile-up backgrounds. Furthermore, precise sub-nanosecond timing serves as a vital supplementary parameter for separating different event topologies and aiding in particle identification via Time-of-Flight (ToF) techniques.

\begin{figure}[htbp]
  \centering
  \includegraphics[width=0.75\textwidth]{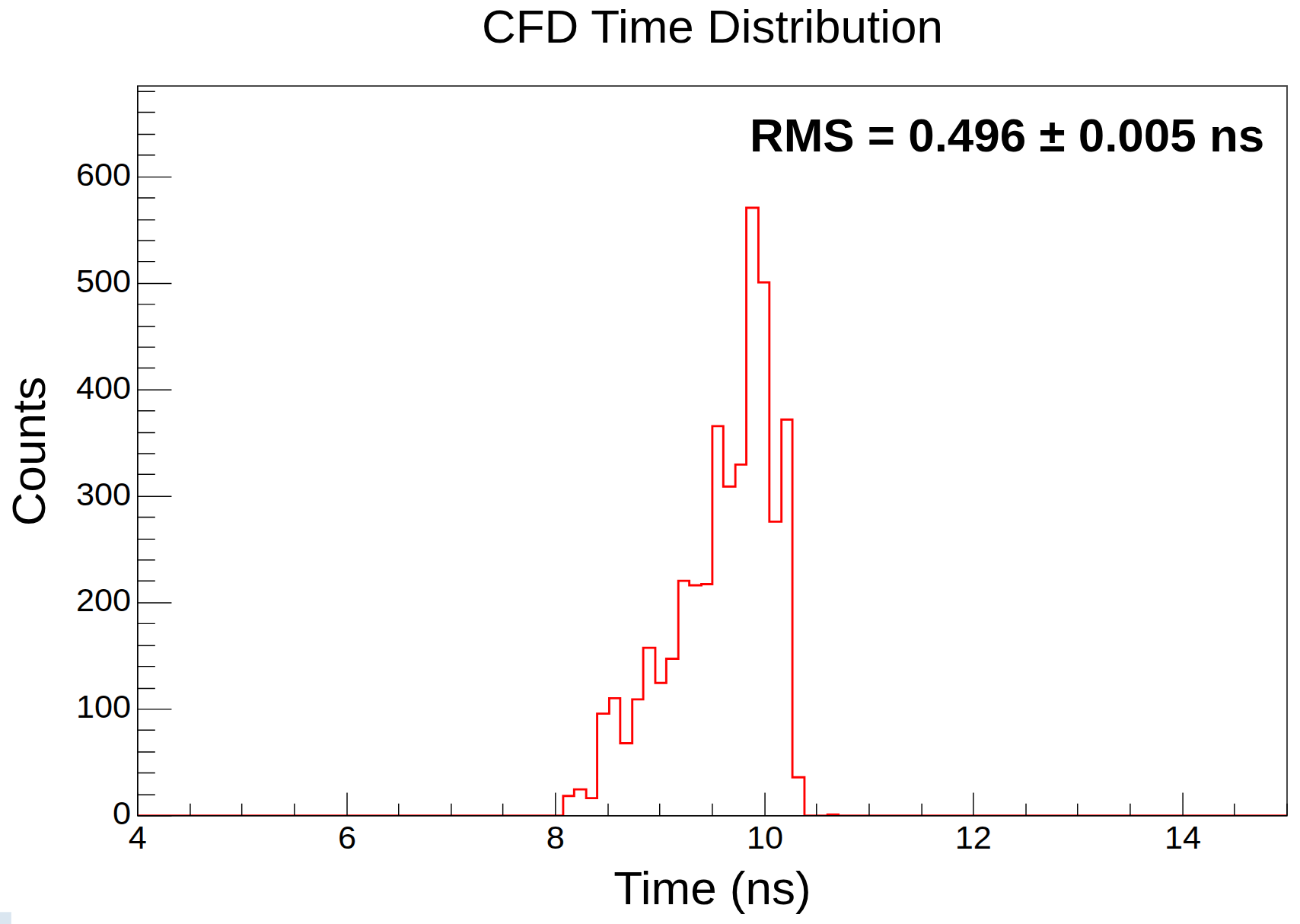}
  \caption{Extracted time resolution of the Spaghetti hadronic calorimeter for 10~GeV incident neutrons, with a Constant Fraction Discriminator (CFD) algorithm at a 10\% threshold. The root-mean-square of the distribution is roughly 497~ps.}
  \label{fig:time_reso}
\end{figure}

The timing performance of the SPACAL was evaluated using the fully digitized PMT waveforms described in Section~\ref{sec:readout_scheme}. To mitigate time-walk effects caused by event-by-event amplitude variations, the reconstructed arrival time is extracted using a Constant Fraction Discriminator (CFD) algorithm~\cite{Harter:2023efw}. Specifically, the trigger time is chosen at a 10\% threshold of the maximum signal amplitude. Fig.~\ref{fig:time_reso} presents the extracted timing resolution distribution for 10~GeV incident neutrons.

A notable feature of this distribution is its distinct asymmetry, characterized by a pronounced tail toward earlier reconstructed arrival times (smaller values on the time axis). This asymmetry is not an artifact of the electronics, but rather a direct physical consequence of different propagation speeds between the incident hadrons and the generated optical photons~\cite{Wigmans:2018fua}. At 10~GeV, incident neutrons travel at highly relativistic velocities ($\beta \approx 1$). This velocity significantly exceeds the local speed of optical photons propagating through the plastic scintillating fibers ($v = c/n \approx 0.63c$, assuming a typical polystyrene core with refractive index of $n \approx 1.58$). 

Because the neutron outpaces its own scintillation light, the precise transverse impact point and the longitudinal depth of the primary hadronic interaction introduce a systematic timing dispersion. Neutrons that happen to enter and travel directly down the scintillating fibers—or those that penetrate deeply into the module before undergoing a hadronic interaction—generate scintillation light physically closer to the rear-mounted photodetectors. These delayed interactions effectively "beat" the light to the back of the calorimeter, yielding systematically earlier reconstructed signal times compared to neutrons that interact promptly near the front face of the dense absorber matrix. 

Despite this inherent structural and kinematic asymmetry, the dense sampling and the fast intrinsic decay time of the plastic scintillators result in a sharply peaked CFD timing distribution. The RMS of the distribution yields a timing resolution of $0.497 \pm 0.004$~ns. This $\sim500$~ps resolution demonstrates that the compact SPACAL provides robust sub-nanosecond timing performance, satisfying the sub-nanosecond timing requirements of the EicC forward physics program.

\subsection{System-Level Performance of Full EMCal+HCal ZDC}
\label{sec:zdc_system_performance}
This subsection presents the system-level performance of the complete ZDC consisting of upstream EMCal and downstream SPACAL HCal, covering transverse position reconstruction and particle identification relying on longitudinal shower morphology across both detector layers.

\subsubsection{Position Reconstruction and Resolution}
\label{sec:position_resolution}
Accurate reconstruction of the transverse position is essential for determining the scattering angle and transverse momentum of forward-going neutral particles. The position resolution is therefore evaluated using the full ZDC, including both the upstream EMCal and downstream hadronic SPACAL. Although hadrons deposit only a relatively small fraction of their total energy in the EMCal, the upstream section can provide valuable spatial information for position reconstruction. In particular, when the first hadronic interaction occurs in the EMCal, the shower is sampled at an earlier stage, before substantial transverse spreading develops. Moreover, the finer transverse granularity of the EMCal provides a more precise measurement of the shower energy distribution than the coarser HCal segmentation. Incorporating the EMCal response can therefore improve the determination of the incident particle position.

\begin{figure}[htbp]
  \centering
  \includegraphics[width=0.60\textwidth]{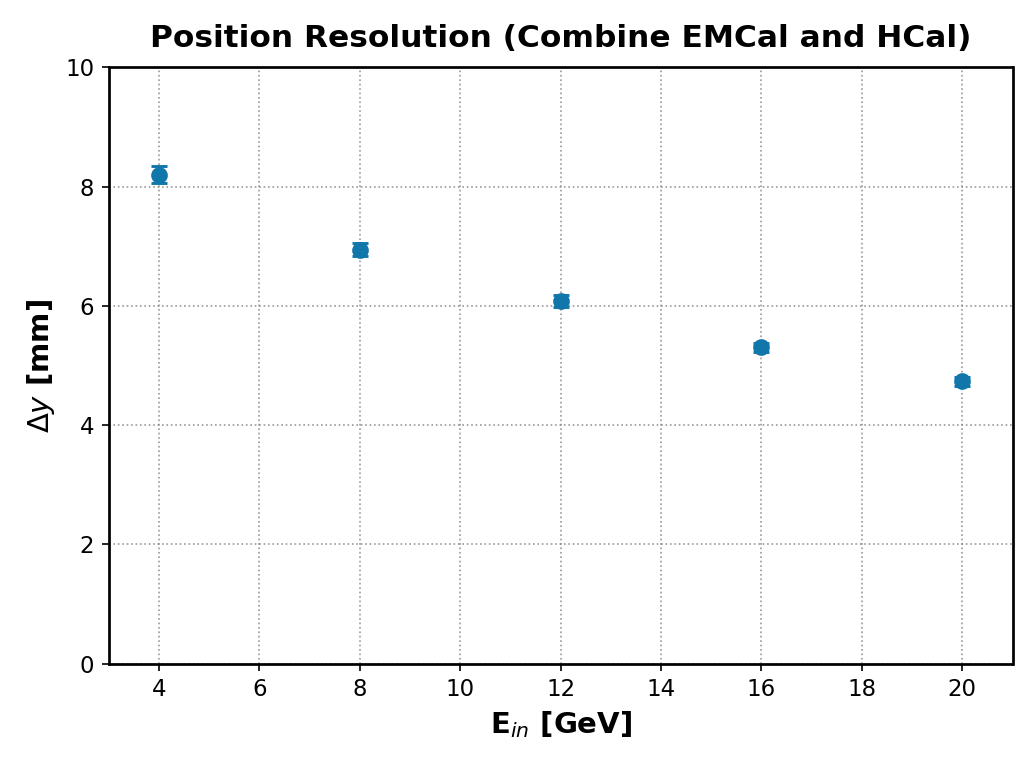}
  \caption{Position resolution of the full EicC ZDC (including both EMCal and HCal sections) as a function of incident neutron energy.}
  \label{ZDC_Pos_Resolution}
\end{figure}

The transverse coordinates $(x_{\mathrm{rec}}, y_{\mathrm{rec}})$ of the incident neutron are reconstructed using a linear energy-weighted center-of-gravity method~\cite{Awes:1992yp}. For each ZDC section, only modules with energy deposition in the sensitive volume above a predefined threshold are considered in the position reconstruction. The thresholds are set to 50~MeV for the $\mathrm{PbWO}_4$ crystal section and 10~MeV for both the Shashlik and HCal sections. Within each section, the module with the largest energy deposition is first identified as the local maximum. A reconstruction window centered on this module is then defined, with a size of $3\times3$ modules for the $\mathrm{PbWO}_4$ and Shashlik sections and $5\times5$ modules for the HCal. A valid position reconstruction is required to contain, in addition to the local-maximum module, at least one other neighboring module within this window whose energy deposition exceeds the corresponding threshold. For a ZDC section satisfying these criteria, the reconstructed transverse coordinates are calculated as
\begin{equation}
x_{\mathrm{rec}}=\frac{\sum_i E_i x_i}{\sum_i E_i},
\qquad
y_{\mathrm{rec}}=\frac{\sum_i E_i y_i}{\sum_i E_i},
\end{equation}
where $E_i$ is the energy deposited in the sensitive volume of the $i$-th module and $(x_i,y_i)$ denotes its transverse center. The sums include the modules within the corresponding reconstruction window that satisfy the energy threshold. The final neutron position is selected according to an upstream-first procedure. If the $\mathrm{PbWO}_4$ section yields a valid reconstructed position, it is adopted as the final neutron position. Otherwise, the Shashlik reconstruction is used when available; if neither upstream EMCal sector provides a valid position, the position reconstructed from the HCal is used.

Fig.~\ref{ZDC_Pos_Resolution} presents the position resolution of the full ZDC system as a function of the incident neutron energy. The resolution improves with increasing neutron energy, as the larger number of secondary particles and the correspondingly more developed transverse shower provide a more stable energy-weighted position estimate. The combination of information from the $\mathrm{PbWO}_4$, Shashlik, and HCal sectors, together with the upstream-first reconstruction strategy, provides sub-centimeter transverse position resolution over the primary EicC kinematic range. Currently, the baseline EMCal Shashlik design provides a highly cost-effective solution when the incident particle multiplicity is low. However, should future physics channels reveal severe cluster overlaps, the system retains the structural flexibility to be upgraded to a high-granularity imaging calorimeter, such as the Tungsten-Silicon (W/Si) calorimeter proposed in the EIC Conceptual Design Report~\cite{AbdulKhalek:2021gbh}.

\subsubsection{Particle Identification}
\label{sec:particle_identification}
Photon--neutron discrimination exploits the distinct longitudinal and transverse shower characteristics of electromagnetic and hadronic particles in the ZDC. All particle-identification studies are performed using the full three-sector configuration. Reliable separation of forward photons from spectator neutrons is essential for isolating physics processes involving forward bremsstrahlung photons and for identifying spectator neutrons in nuclear collisions.

The ZDC is located downstream of a series of dipole and quadrupole magnets, which deflect the ion beam together with most high-energy charged particles produced at the interaction point, away from the ZDC acceptance. The residual charged-particle background can be further suppressed using a veto counter placed immediately upstream of the ZDC. The remaining particle-identification problem is dominated by the separation of photons and neutrons using their shower topology in the calorimeter system. Electromagnetic showers initiated by photons are characterized by the radiation length, $X_{0}$, and a relatively small Moli\`ere radius. Consequently, photons generally produce compact and localized energy depositions, with a large fraction of the shower contained in the electromagnetic calorimeters. In contrast, neutron-induced hadronic showers are governed by the substantially larger nuclear interaction length $\lambda_{\mathrm{I}}$. Neutrons may produce only sparse preshower activity in the upstream detectors and deposit a larger fraction of their energy in the downstream hadronic calorimeter, typically with a broader transverse profile and larger event-by-event fluctuations.

A custom transformer-based classifier, developed in this work and hereafter referred to as the ZDC Transformer, is employed to exploit these complementary shower signatures. For each event, the energy depositions measured by the PbWO$_4$ crystal calorimeter, the Shashlik electromagnetic calorimeter, and the SPACAL hadronic calorimeter are used as three input channels. Depositions below the analysis threshold are set to zero, and the detector response of each event is normalized by its maximum deposited energy. This preprocessing reduces the direct sensitivity of the classifier to the overall energy scale while preserving the spatial and longitudinal differences between electromagnetic and hadronic shower development.

The detector responses are converted into embedded shower tokens using a convolutional patch-embedding layer with learnable positional encoding. The resulting token sequence is processed by a transformer-based classifier consisting of four encoder layers with multi-head self-attention, followed by feature aggregation and a fully connected classification head. The model contains approximately $1.65\times10^{5}$ trainable parameters.

The simulated data set includes photon and neutron events at nine incident energies from $3$ to $19~\mathrm{GeV}$ in steps of $2~\mathrm{GeV}$. At each energy, $50,000$ events of each particle type are used for model development and divided into training and validation samples using a stratified 80:20 split. An independent test sample of $5,000$ photon and $5,000$ neutron events per energy is reserved for performance evaluation. The network is trained using the AdamW optimizer with binary cross-entropy loss, and the model with the lowest validation loss is retained for the final evaluation.

\begin{figure}[htbp]
    \centering
    \includegraphics[width=0.60\textwidth]{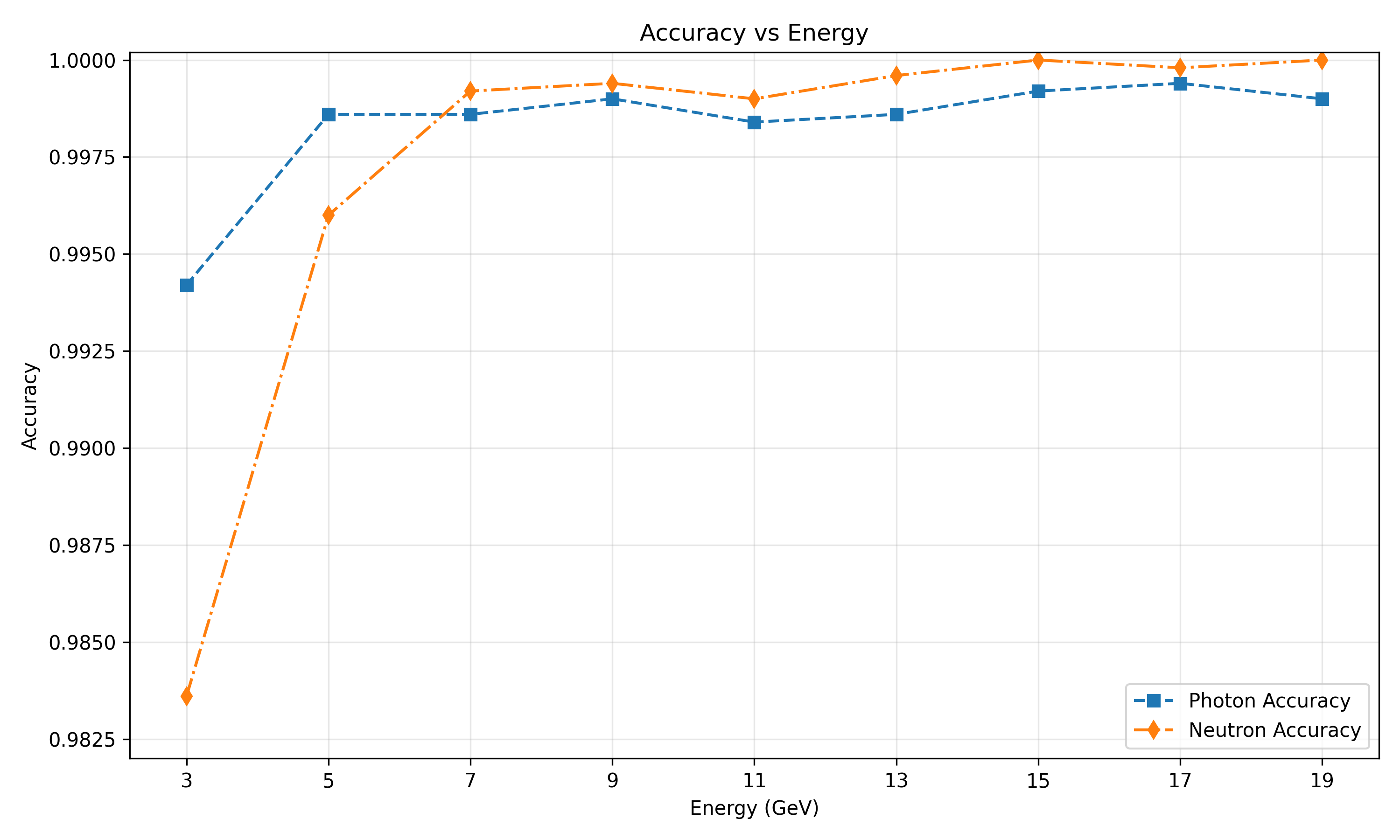}
    \caption{
        Photon--neutron identification performance of the ZDC Transformer as
        a function of incident energy. Each point is evaluated using $5\,000$
        photon events and $5\,000$ neutron events from the held-out test sample.
    }
    \label{fig:PID}
\end{figure}

Figure~\ref{fig:PID} shows the photon--neutron identification performance as a function of incident energy. At $3~\mathrm{GeV}$, the overall classification accuracy is approximately $98.9\%$, while the photon and neutron correct-identification rates are approximately $99.4\%$ and $98.4\%$, respectively. The lower neutron correct-identification rate at this energy is consistent with the relatively large fluctuations and lower secondary-particle multiplicity of low-energy hadronic showers. As the incident energy increases, the electromagnetic and hadronic shower signatures become more distinct, providing richer spatial and longitudinal information to the classifier. The overall accuracy exceeds approximately $99.8\%$ for incident energies of $7~\mathrm{GeV}$ and above and approaches $99.9\%$ at the highest energies considered.

When all nine energy points are combined, the classifier achieves an overall test accuracy of approximately $99.78\%$. Within the simulated samples considered in this study, these results demonstrate that the transformer classifier effectively combines the energy-deposition information from the three detector subsystems to separate photon- and neutron-induced showers. It is worth noting that the present study uses only the spatial energy-deposition information of the detector modules; timing information is not included in the classifier input. Incorporating the timing response in future studies is expected to provide additional discrimination power and further improve the particle-identification performance.

\section{Summary and Outlook}
\label{sec:summary_and_outlook}

In this paper, we have presented the conceptual design and comprehensive Geant4 simulation study of the hadronic section for the EicC Zero-Degree Calorimeter (ZDC). Driven by stringent spatial constraints and demanding forward physics requirements, a Spaghetti Calorimeter (SPACAL) architecture was established as the optimal baseline. The finalized geometry features a $12 \times 12$ array of $5 \times 5$ cm$^2$ modules with a longitudinal depth of 110 cm. The active medium utilizes single-clad plastic scintillating fibers embedded within a dense copper capillary tube matrix. This highly compact configuration allows the fiber bundles to be directly coupled to PMT arrays, seamlessly integrating into the restricted EicC interaction region while completely eliminating the need for bulky adiabatic light guides.

An end-to-end digitization simulation framework integrating realistic optical transport, measured PMT single-photon responses, and front-end electronic noise is established to evaluate the detector physics performance. It achieves an energy resolution of 11.9\% and time resolution of 500~ps for typical 10 GeV incident neutrons. An energy scan for neutrons at 6–20 GeV validates that the baseline design meets the core performance requirements, achieving a stochastic term of $a = (33.42 \pm 0.86)\%$ and a constant term of $b = (1.47 \pm 0.24)\%$. Benefiting from fine transverse segmentation, a sub-centimeter position resolution is realized when combining with the upstream EMCal. Moreover, full-system shower shape analysis leveraging the complementary response of SPACAL and EMCal enables robust particle identification with an accuracy over 99\% for neutron-photon separation in most of the kinematic region.

Looking forward, the inherent structural flexibility of the SPACAL architecture provides a natural pathway for advanced performance upgrades. Most notably, the design is highly compatible with dual-readout calorimetry~\cite{CHO2025100021,RevModPhys.90.025002,wigmans2000calorimetry,
AKCHURIN2014130,LEE201776,ANTONELLO201852,ANTONELLO2019127}. By substituting a fraction of the scintillating fibers with clear quartz fibers, both scintillation and Cherenkov signals could be simultaneously extracted from the same module. This would enable event-by-event corrections of the electromagnetic fraction within hadronic showers, mitigating non-compensation effects and significantly improving the ultimate hadronic energy resolution without requiring sweeping mechanical redesigns~\cite{Akchurin:2005an}. 

The immediate next steps for the EicC ZDC project involve transitioning from simulation to hardware. This includes the engineering design and construction of a reduced-scale mechanical prototype, followed by dedicated beam tests. These experimental campaigns will be crucial for validating the simulated optical efficiencies, calibrating the front-end electronics chain, and verifying the expected detector performance in a realistic physical environment, ultimately laying the groundwork for the final EicC forward detector construction.

\acknowledgments
We would like to express our gratitude to Jianing Dong for ensuring the performance and reliability of the computer facility at Shandong University.

This work is supported by the National Key Research and Development Project of China under Contract No.~2023YFA1606800, and National Natural Science Foundation of China under Grant Nos.~12305144 and 12375139, and the Shandong Province Natural Science Foundation under Grant No.~2023HWYQ-010.


\bibliographystyle{JHEP}
\bibliography{biblio.bib}

@article{Anderle:2021wcy,
    author = "Anderle, Daniele P. and others",
    title = "{Electron-ion collider in China}",
    eprint = "2102.09222",
    archivePrefix = "arXiv",
    primaryClass = "nucl-ex",
    reportNumber = "Frontiers of Physics, Volume 16 Issue (6):64701, 2021",
    doi = "10.1007/s11467-021-1062-0",
    journal = "Front. Phys. (Beijing)",
    volume = "16",
    number = "6",
    pages = "64701",
    year = "2021"
}

@article{GEANT4:2002zbu,
    author = "Agostinelli, S. and others",
    collaboration = "GEANT4",
    title = "{GEANT4 - A Simulation Toolkit}",
    reportNumber = "SLAC-PUB-9350, FERMILAB-PUB-03-339, CERN-IT-2002-003",
    doi = "10.1016/S0168-9002(03)01368-8",
    journal = "Nucl. Instrum. Meth. A",
    volume = "506",
    pages = "250--303",
    year = "2003"
}

@article{Armstrong:1998qs,
    author = "Armstrong, T. A and others",
    title = "{The E864 lead-scintillating fiber hadronic calorimeter}",
    doi = "10.1016/S0168-9002(98)91984-2",
    journal = "Nucl. Instrum. Meth. A",
    volume = "406",
    pages = "227--258",
    year = "1998"
}

@article{AKCHURIN2014130,
title = {The electromagnetic performance of the RD52 fiber calorimeter},
journal = {Nuclear Instruments and Methods in Physics Research Section A: Accelerators, Spectrometers, Detectors and Associated Equipment},
volume = {735},
pages = {130-144},
year = {2014},
issn = {0168-9002},
doi = {https://doi.org/10.1016/j.nima.2013.09.033},
url = {https://www.sciencedirect.com/science/article/pii/S0168900213012679},
author = {N. Akchurin and F. Bedeschi and A. Cardini and M. Cascella and F. Cei and D. {De Pedis} and R. Ferrari and S. Fracchia and S. Franchino and M. Fraternali and G. Gaudio and P. Genova and J. Hauptman and L. {La Rotonda} and S. Lee and M. Livan and E. Meoni and A. Moggi and D. Pinci and A. Policicchio and J.G. Saraiva and F. Scuri and A. Sill and T. Venturelli and R. Wigmans}
}

@article{LEE201776,
title = {Hadron detection with a dual-readout fiber calorimeter},
journal = {Nuclear Instruments and Methods in Physics Research Section A: Accelerators, Spectrometers, Detectors and Associated Equipment},
volume = {866},
pages = {76-90},
year = {2017},
issn = {0168-9002},
doi = {https://doi.org/10.1016/j.nima.2017.05.025},
url = {https://www.sciencedirect.com/science/article/pii/S0168900217305703},
author = {S. Lee and A. Cardini and M. Cascella and S. Choi and G. Ciapetti and R. Ferrari and S. Franchino and M. Fraternali and G. Gaudio and S. Ha and J. Hauptman and H. Kim and A. Lanza and F. Li and M. Livan and E. Meoni and J. Park and F. Scuri and A. Sill and R. Wigmans}
}

@article{CARDINI201641,
title = {The small-angle performance of a dual-readout fiber calorimeter},
journal = {Nuclear Instruments and Methods in Physics Research Section A: Accelerators, Spectrometers, Detectors and Associated Equipment},
volume = {808},
pages = {41-53},
year = {2016},
issn = {0168-9002},
doi = {https://doi.org/10.1016/j.nima.2015.11.005},
url = {https://www.sciencedirect.com/science/article/pii/S0168900215013674},
author = {A. Cardini and M. Cascella and S. Choi and D. {De Pedis} and R. Ferrari and S. Franchino and G. Gaudio and S. Ha and J. Hauptman and L. {La Rotonda} and S. Lee and F. Li and M. Livan and E. Meoni and F. Scuri and A. Sill and R. Wigmans}
}

@article{ANTONELLO201852,
title = {Tests of a dual-readout fiber calorimeter with SiPM light sensors},
journal = {Nuclear Instruments and Methods in Physics Research Section A: Accelerators, Spectrometers, Detectors and Associated Equipment},
volume = {899},
pages = {52-64},
year = {2018},
issn = {0168-9002},
doi = {https://doi.org/10.1016/j.nima.2018.05.016},
url = {https://www.sciencedirect.com/science/article/pii/S0168900218306077},
author = {M. Antonello and M. Caccia and M. Cascella and M. Dunser and R. Ferrari and S. Franchino and G. Gaudio and K. Hall and J. Hauptman and H. Jo and K. Kang and B. Kim and S. Lee and G. Lerner and L. Pezzotti and R. Santoro and I. Vivarelli and R. Ye and R. Wigmans}
}

@article{ANTONELLO2019127,
title = {Development of a Silicon Photomultiplier based dual readout calorimeter: The pathway beyond the proof-of-concept},
journal = {Nuclear Instruments and Methods in Physics Research Section A: Accelerators, Spectrometers, Detectors and Associated Equipment},
volume = {936},
pages = {127-129},
year = {2019},
note = {Frontier Detectors for Frontier Physics: 14th Pisa Meeting on Advanced Detectors},
issn = {0168-9002},
doi = {https://doi.org/10.1016/j.nima.2018.10.169},
url = {https://www.sciencedirect.com/science/article/pii/S0168900218314839},
author = {M. Antonello and M. Caccia and R. Ferrari and L. Pezzotti and R. Santoro}
}

@article{Lee:2017xss,
    author = "Lee, Sehwook and Livan, Michele and Wigmans, Richard",
    title = "{Dual-Readout Calorimetry}",
    eprint = "1712.05494",
    archivePrefix = "arXiv",
    primaryClass = "physics.ins-det",
    doi = "10.1103/RevModPhys.90.025002",
    journal = "Rev. Mod. Phys.",
    volume = "90",
    number = "2",
    pages = "025002",
    year = "2018"
}

@article{Ji:2004gf,
    author = "Ji, X.",
    title = "{Generalized parton distributions}",
    doi = "10.1146/annurev.nucl.54.070103.181302",
    journal = "Ann. Rev. Nucl. Part. Sci.",
    volume = "54",
    pages = "413--450",
    year = "2004"
}

@article{Chakrabarti:2005zm,
    author = "Chakrabarti, D. and Mukherjee, A.",
    title = "{Generalized parton distributions in the impact parameter space with non-zero skewedness}",
    eprint = "hep-ph/0506006",
    archivePrefix = "arXiv",
    doi = "10.1103/PhysRevD.72.034013",
    journal = "Phys. Rev. D",
    volume = "72",
    pages = "034013",
    year = "2005"
}

@article{Scopetta:2003et,
    author = "Scopetta, Sergio and Vento, Vicente",
    title = "{Generalized parton distributions and composite constituent quarks}",
    eprint = "hep-ph/0307150",
    archivePrefix = "arXiv",
    reportNumber = "FTUV-03-0708, IFIC-03-35",
    doi = "10.1103/PhysRevD.69.094004",
    journal = "Phys. Rev. D",
    volume = "69",
    pages = "094004",
    year = "2004"
}

@article{Roberts:2021nhw,
    author = "Roberts, Craig D. and Richards, David G. and Horn, Tanja and Chang, Lei",
    title = "{Insights into the emergence of mass from studies of pion and kaon structure}",
    eprint = "2102.01765",
    archivePrefix = "arXiv",
    primaryClass = "hep-ph",
    reportNumber = "NJU-INP 034/21",
    doi = "10.1016/j.ppnp.2021.103883",
    journal = "Prog. Part. Nucl. Phys.",
    volume = "120",
    pages = "103883",
    year = "2021"
}

@article{Ding:2022ows,
    author = "Ding, Minghui and Roberts, Craig D. and Schmidt, Sebastian M.",
    title = "{Emergence of Hadron Mass and Structure}",
    eprint = "2211.07763",
    archivePrefix = "arXiv",
    primaryClass = "hep-ph",
    reportNumber = "NJU-INP 066/22",
    doi = "10.3390/particles6010004",
    journal = "Particles",
    volume = "6",
    number = "1",
    pages = "57--120",
    year = "2023"
}

@article{Arrington:2021biu,
    author = "Arrington, J. and others",
    title = "{Revealing the structure of light pseudoscalar mesons at the electron{\textendash}ion collider}",
    eprint = "2102.11788",
    archivePrefix = "arXiv",
    primaryClass = "nucl-ex",
    doi = "10.1088/1361-6471/abf5c3",
    journal = "J. Phys. G",
    volume = "48",
    number = "7",
    pages = "075106",
    year = "2021"
}

@article{Guo:2017jvc,
    author = "Guo, Feng-Kun and Hanhart, Christoph and Mei{\ss}ner, Ulf-G. and Wang, Qian and Zhao, Qiang and Zou, Bing-Song",
    title = "{Hadronic molecules}",
    eprint = "1705.00141",
    archivePrefix = "arXiv",
    primaryClass = "hep-ph",
    doi = "10.1103/RevModPhys.90.015004",
    journal = "Rev. Mod. Phys.",
    volume = "90",
    number = "1",
    pages = "015004",
    year = "2018",
    note = "[Erratum: Rev.Mod.Phys. 94, 029901 (2022)]"
}

@article{Brambilla:2019esw,
    author = "Brambilla, Nora and Eidelman, Simon and Hanhart, Christoph and Nefediev, Alexey and Shen, Cheng-Ping and Thomas, Christopher E. and Vairo, Antonio and Yuan, Chang-Zheng",
    title = "{The $XYZ$ states: experimental and theoretical status and perspectives}",
    eprint = "1907.07583",
    archivePrefix = "arXiv",
    primaryClass = "hep-ex",
    reportNumber = "TUM-EFT 125/19",
    doi = "10.1016/j.physrep.2020.05.001",
    journal = "Phys. Rept.",
    volume = "873",
    pages = "1--154",
    year = "2020"
}

@article{Boussarie:2023izj,
    author = "Boussarie, Renaud and others",
    title = "{TMD Handbook}",
    eprint = "2304.03302",
    archivePrefix = "arXiv",
    primaryClass = "hep-ph",
    reportNumber = "JLAB-THY-23-3780, LA-UR-21-20798, MIT-CTP/5386",
    month = "4",
    year = "2023"
}

@article{Angeles-Martinez:2015sea,
    author = "Angeles-Martinez, R. and others",
    title = "{Transverse Momentum Dependent (TMD) parton distribution functions: status and prospects}",
    eprint = "1507.05267",
    archivePrefix = "arXiv",
    primaryClass = "hep-ph",
    doi = "10.5506/APhysPolB.46.2501",
    journal = "Acta Phys. Polon. B",
    volume = "46",
    number = "12",
    pages = "2501--2534",
    year = "2015"
}

@article{Chen:2016qju,
    author = "Chen, Hua-Xing and Chen, Wei and Liu, Xiang and Zhu, Shi-Lin",
    title = "{The hidden-charm pentaquark and tetraquark states}",
    eprint = "1601.02092",
    archivePrefix = "arXiv",
    primaryClass = "hep-ph",
    doi = "10.1016/j.physrep.2016.05.004",
    journal = "Phys. Rept.",
    volume = "639",
    pages = "1--121",
    year = "2016"
}

@article{Favart:2015umi,
    author = "Favart, L. and Guidal, M. and Horn, T. and Kroll, P.",
    title = "{Deeply Virtual Meson Production on the nucleon}",
    eprint = "1511.04535",
    archivePrefix = "arXiv",
    primaryClass = "hep-ph",
    doi = "10.1140/epja/i2016-16158-6",
    journal = "Eur. Phys. J. A",
    volume = "52",
    number = "6",
    pages = "158",
    year = "2016"
}

@article{Sullivan:1971kd,
    author = "Sullivan, J. D.",
    title = "{One pion exchange and deep inelastic electron - nucleon scattering}",
    doi = "10.1103/PhysRevD.5.1732",
    journal = "Phys. Rev. D",
    volume = "5",
    pages = "1732--1737",
    year = "1972"
}

@article{AbdulKhalek:2021gbh,
    author = "Abdul Khalek, R. and others",
    title = "{Science Requirements and Detector Concepts for the Electron-Ion Collider: EIC Yellow Report}",
    eprint = "2103.05419",
    archivePrefix = "arXiv",
    primaryClass = "physics.ins-det",
    reportNumber = "BNL-220990-2021-FORE, JLAB-PHY-21-3198, LA-UR-21-20953",
    doi = "10.1016/j.nuclphysa.2022.122447",
    journal = "Nucl. Phys. A",
    volume = "1026",
    pages = "122447",
    year = "2022"
}

@article{Lu:2025bnm,
    author = "Lu, Zongyang and Yu, Zihan and Lin, Ting and Liang, Yu-Tie and Wang, Rong and Chang, Wan and Xiong, Weizhi",
    title = "{Feasibility study of pion and kaon structure via the Sullivan process at the EicC}",
    eprint = "2512.01720",
    archivePrefix = "arXiv",
    primaryClass = "hep-ex",
    doi = "10.1103/1s8f-9cny",
    journal = "Phys. Rev. D",
    volume = "113",
    number = "11",
    pages = "114002",
    year = "2026"
}

@article{Awes:1992yp,
    author = "Awes, T. C. and Obenshain, F. E. and Plasil, F. and Saini, S. and Sorensen, S. P. and Young, G. R.",
    title = "{A Simple method of shower localization and identification in laterally segmented calorimeters}",
    doi = "10.1016/0168-9002(92)90858-2",
    journal = "Nucl. Instrum. Meth. A",
    volume = "311",
    pages = "130--138",
    year = "1992"
}

@article{Akchurin:2005an,
    author = "Akchurin, N. and Carrell, K. and Hauptman, J. and Kim, H. and Paar, H. P. and Penzo, A. and Thomas, R. and Wigmans, R.",
    title = "{Hadron and jet detection with a dual-readout calorimeter}",
    doi = "10.1016/j.nima.2004.07.285",
    journal = "Nucl. Instrum. Meth. A",
    volume = "537",
    pages = "537--561",
    year = "2005"
}

@article{Losekamm:2024yvw,
    author = {Losekamm, Martin J. and Paul, Stephan and P{\"o}schl, Thomas},
    title = "{Position-dependent light yield in short, coated SCSF-78 scintillating fibers}",
    doi = "10.1016/j.radmeas.2024.107116",
    journal = "Radiat. Meas.",
    volume = "174",
    pages = "107116",
    year = "2024"
}

@article{Bravar:2022xbz,
    author = "Bravar, A. and Demets, Y.",
    title = "{Timing properties of blue-emitting scintillating fibers}",
    eprint = "2203.13322",
    archivePrefix = "arXiv",
    primaryClass = "physics.ins-det",
    doi = "10.1088/1748-0221/17/12/P12020",
    journal = "JINST",
    volume = "17",
    number = "12",
    pages = "P12020",
    year = "2022"
}

@article{Mezrag:2023nkp,
    author = "Mezrag, C{\'e}dric",
    title = "{Generalised Parton Distributions in Continuum Schwinger Methods: Progresses, Opportunities and Challenges}",
    doi = "10.3390/particles6010015",
    journal = "Particles",
    volume = "6",
    number = "1",
    pages = "262--296",
    year = "2023"
}

@article{Qiu:2024pvw,
    author = "Qiu, Jian-Wei and Yu, Zhite",
    title = "{Overview on the theory and phenomenology of generalized parton distributions}",
    eprint = "2410.17432",
    archivePrefix = "arXiv",
    primaryClass = "hep-ph",
    reportNumber = "JLAB-THY-24-4223",
    doi = "10.22323/1.477.0035",
    journal = "PoS",
    volume = "Transversity2024",
    pages = "035",
    year = "2024"
}

@article{Zheng:2014cha,
    author = "Zheng, L. and Aschenauer, E. C. and Lee, J. H.",
    title = "{Determination of electron-nucleus collision geometry with forward neutrons}",
    eprint = "1407.8055",
    archivePrefix = "arXiv",
    primaryClass = "hep-ex",
    doi = "10.1140/epja/i2014-14189-3",
    journal = "Eur. Phys. J. A",
    volume = "50",
    number = "12",
    pages = "189",
    year = "2014"
}

@article{Hegazy:2024tgt,
    author = "Hegazy, Mariam and Rafaat, Aliaa and Magdy, Niseem and Li, Wenliang and Deshpande, Abhay and Abdelhady, A. M. H. and Ellithi, A. Y.",
    title = "{Centrality definition in e + A collisions at the electron{\textendash}ion collider}",
    eprint = "2411.07963",
    archivePrefix = "arXiv",
    primaryClass = "hep-ph",
    doi = "10.1088/1361-6471/ad9344",
    journal = "J. Phys. G",
    volume = "52",
    number = "1",
    pages = "015002",
    year = "2025"
}

@article{Simeonov:2022yqy,
    author = "Simeonov, Radoslav",
    collaboration = "ALICE",
    title = "{Design and Test-Beam Results of the FoCal-H Demonstrator Prototype}",
    eprint = "2211.14791",
    archivePrefix = "arXiv",
    primaryClass = "physics.ins-det",
    doi = "10.3390/instruments6040070",
    journal = "Instruments",
    volume = "6",
    number = "4",
    pages = "70",
    year = "2022"
}

@article{Acosta:1990qs,
    author = "Acosta, D. and others",
    title = "{Results of Prototype Studies for a Spaghetti Calorimeter}",
    reportNumber = "CERN-EP/90-37, CERN-LAA-HC-90-004",
    doi = "10.1016/0168-9002(90)91833-W",
    journal = "Nucl. Instrum. Meth. A",
    volume = "294",
    pages = "193--210",
    year = "1990"
}

@article{Acosta1992LateralSP,
  title={Lateral shower profiles in a lead/scintillating fiber calorimeter},
  author={Darin Acosta and Salvatore Buontempo and Luiz Pereira Cal{\^o}ba and Riccardo DeSalvo and Antonio Ereditato and Roberto Ferrari and G. Fumagalli and G. Goggi and W. Hao and Ana. Henriques and L. Linssen and Ma Liyan and Am{\'e}lia Maio and Maria Rosa Mondardini and B. Ong and H. P. Paar and Francesca Pastore and E. Pennacchio and Luc Poggioli and Giacomo Polesello and Flavio Riccardi and Adele Rimoldi and C. V. Scheel and J. M. Seixas and A. Sim{\'o}n and M. Sivertz and Peter Sonderegger and M. N. Souza and Zieli Dutra Thome and Valerio Vercesi and Yibin Wang and Richard Wigmans and C. Xu},
  journal={Nuclear Instruments \& Methods in Physics Research Section A-accelerators Spectrometers Detectors and Associated Equipment},
  year={1992},
  volume={316},
  pages={184-201},
  url={https://api.semanticscholar.org/CorpusID:121634492}
}

@article{ADINOLFI2002326,
title = {The KLOE electromagnetic calorimeter},
journal = {Nuclear Instruments and Methods in Physics Research Section A: Accelerators, Spectrometers, Detectors and Associated Equipment},
volume = {494},
number = {1},
pages = {326-331},
year = {2002},
note = {Proceedings of the 8th International Conference on Instrumentatio n for Colliding Beam Physics},
issn = {0168-9002},
doi = {https://doi.org/10.1016/S0168-9002(02)01488-2},
url = {https://www.sciencedirect.com/science/article/pii/S0168900202014882},
author = {M. Adinolfi and F. Ambrosino and A. Antonelli and M. Antonelli and F. Anulli and G. Barbiellini and G. Bencivenni and S. Bertolucci and C. Bini and C. Bloise and V. Bocci and F. Bossi and P. Branchini and G. Cabibbo and R. Caloi and P. Campana and M. Casarsa and G. Cataldi and F. Ceradini and F. Cervelli and P. Ciambrone and E. {De Lucia} and P. {De Simone} and G. {De Zorzi} and S. Dell'Agnello and A. Denig and A. {Di Domenico} and C. {Di Donato} and S. {Di Falco} and A. Doria and O. Erriquez and A. Farilla and A. Ferrari and M.L. Ferrer and G. Finocchiaro and C. Forti and A. Franceschi and P. Franzini and M.L. Gao and C. Gatti and P. Gauzzi and A. Giannasi and S. Giovannella and E. Graziani and H.G. Han and S.W. Han and X. Huang and M. Incagli and L. Ingrosso and L. Keeble and W. Kim and C. Kuo and G. Lanfranchi and J. Lee-Franzini and T. Lomtadze and C.S. Mao and M. Martemianov and W. Mei and R. Messi and S. Miscetti and S. Moccia and M. Moulson and S. Müller and F. Murtas and L. Pacciani and M. Palomba and M. Palutan and E. Pasqualucci and L. Passalacqua and A. Passeri and D. Picca and G. Pirozzi and L. Pontecorvo and M. Primavera and P. Santangelo and E. Santovetti and G. Saracino and R.D. Schamberger and B. Sciascia and F. Scuri and I. Sfiligoi and P. Silano and T. Spadaro and E. Spiriti and L. Tortora and P. Valente and B. Valeriani and G. Venanzoni and A. Ventura and S. Wölfle and Y. Wu and Y.G. Xie and P.F. Zema and C.D. Zhang and J.Q. Zhang and P.P. Zhao}
}

@article{friend1976measurements,
  title={Measurements of energy flow distributions of 10 GeV/c hadronic showers in iron and in aluminium},
  author={Friend, B. and King, A. and Kiss, D. and Schmidt-Parzefall, W. and Winter, K. and Niebergall, F. and Wilmsen, W.},
  journal={Nuclear Instruments and Methods},
  volume={136},
  number={3},
  pages={505--510},
  year={1976},
  doi={10.1016/0029-554X(76)90370-0}
}

@article{HOLDER197869,
title = {Performance of a magnetized total absorption calorimeter between 15 GeV and 140 GeV},
journal = {Nuclear Instruments and Methods},
volume = {151},
number = {1},
pages = {69-80},
year = {1978},
issn = {0029-554X},
doi = {https://doi.org/10.1016/0029-554X(78)90472-X},
url = {https://www.sciencedirect.com/science/article/pii/0029554X7890472X},
author = {M. Holder and J. Knobloch and J. May and H.P. Paar and P. Palazzi and D. Schlatter and J. Steinberger and H. Suter and H. Wahl and E.G.H. Williams and F. Eisele and C. Geweniger and K. Kleinknecht and G. Spahn and H.-J. Willutzki and W. Dorth and F. Dydak and T. Flottmann and V. Hepp and K. Tittel and J. Wotschack and P. Bloch and B. Devaux and M. Grimm and J. Maillard and B. Peyaud and J. Rander and A. Savoy-Navarro and R. Turlay and F.L. Navarria}
}

@article{Cheshire1975MeasurementsOT,
  title={Measurements of the development of cascades in a tungsten-scintillator ionization spectrometer},
  author={D. L. Cheshire and R. W. Huggett and D. P. Johnson and William Vernon Jones and Steven P. Rountree and Wolfgang K. H. Schmidt and Richard J. Kurz and T. C. Bowen and D. A. Delise and Edmund Philip Krider and Charles D. Orth},
  journal={Nuclear Instruments and Methods},
  year={1975},
  volume={126},
  pages={253-262},
  url={https://api.semanticscholar.org/CorpusID:120243627}
}

@article{PhysRevD.12.2587,
  title = {Inelastic-interaction mean free path of negative pions in tungsten},
  author = {Cheshire, D. L. and Huggett, R. W. and Jones, W. V. and Rountree, S. P. and Schmidt, W. K. H. and Kurz, R. J. and Bowen, T. and DeLise, D. A. and Krider, E. P. and Orth, C. D.},
  journal = {Phys. Rev. D},
  volume = {12},
  issue = {9},
  pages = {2587--2593},
  numpages = {0},
  year = {1975},
  month = {Nov},
  publisher = {American Physical Society},
  doi = {10.1103/PhysRevD.12.2587},
  url = {https://link.aps.org/doi/10.1103/PhysRevD.12.2587}
}

@article{GRANT1975167,
title = {A Monte Carlo calculation of high energy hadronic cascade in matter},
journal = {Nuclear Instruments and Methods},
volume = {131},
number = {1},
pages = {167-172},
year = {1975},
issn = {0029-554X},
doi = {https://doi.org/10.1016/0029-554X(75)90372-9},
url = {https://www.sciencedirect.com/science/article/pii/0029554X75903729},
author = {A. Grant}
}

@article{WIGMANS1987389,
title = {On the energy resolution of uranium and other hadron calorimeters},
journal = {Nuclear Instruments and Methods in Physics Research Section A: Accelerators, Spectrometers, Detectors and Associated Equipment},
volume = {259},
number = {3},
pages = {389-429},
year = {1987},
issn = {0168-9002},
doi = {https://doi.org/10.1016/0168-9002(87)90823-0},
url = {https://www.sciencedirect.com/science/article/pii/0168900287908230},
author = {Richard Wigmans}
}

@article{Speckmayer:1443830,
      author        = "Speckmayer, P and Grefe, C",
      title         = "{Comparison of the Performance of Tungsten and Steel
                       Hadronic Sampling Calorimeters}",
      year          = "2012",
      url           = "https://cds.cern.ch/record/1443830",
}

@article{ACOSTA1990193,
title = {Results of prototype studies for a spaghetti calorimeter},
journal = {Nuclear Instruments and Methods in Physics Research Section A: Accelerators, Spectrometers, Detectors and Associated Equipment},
volume = {294},
number = {1},
pages = {193-210},
year = {1990},
issn = {0168-9002},
doi = {https://doi.org/10.1016/0168-9002(90)91833-W},
url = {https://www.sciencedirect.com/science/article/pii/016890029091833W},
author = {D. Acosta and S. Buontempo and L. Calôba and M. Caria and R. Desalvo and A. Ereditato and R. Ferrari and M. Fraternali and G. Fumagalli and O. Gildemeister and F.G. Hartjes and Th.H. Henkes and A. Henriques and L. Linssen and M. Livan and A. Maio and L. Mapelli and K.H. Meier and B. Ong and H.P. Paar and F. Pastore and M. Pereira and L. Poggioli and C.V. Scheel and J.M. Seixas and A. Simon and M. Sivertz and P. Sonderegger and M.N. Souza and Z.D. Thomé and V. Vercesi and R. Wigmans}
}

@article{DESALVO1995122,
title = {Lead/scintillating fibre (“Spaghetti”) calorimetry},
journal = {Nuclear Physics B - Proceedings Supplements},
volume = {44},
number = {1},
pages = {122-131},
year = {1995},
issn = {0920-5632},
doi = {https://doi.org/10.1016/S0920-5632(95)80018-2},
url = {https://www.sciencedirect.com/science/article/pii/S0920563295800182},
author = {R. DeSalvo}
}

@article{BRAVAR2024168766,
title = {Development of the scintillating fiber timing detector for the Mu3e experiment},
journal = {Nuclear Instruments and Methods in Physics Research Section A: Accelerators, Spectrometers, Detectors and Associated Equipment},
volume = {1058},
pages = {168766},
year = {2024},
issn = {0168-9002},
doi = {https://doi.org/10.1016/j.nima.2023.168766},
url = {https://www.sciencedirect.com/science/article/pii/S016890022300757X},
author = {A. Bravar and A. Buonaura and S. Corrodi and A. Damyanova and Y. Demets and L. Gerritzen and C. Grab and C. Martin Perez and A. Papa}
}

@article{LI2026171595,
title = {Performance evaluation of compact plastic scintillating fiber modules for muon tomography applications},
journal = {Nuclear Instruments and Methods in Physics Research Section A: Accelerators, Spectrometers, Detectors and Associated Equipment},
volume = {1089},
pages = {171595},
year = {2026},
issn = {0168-9002},
doi = {https://doi.org/10.1016/j.nima.2026.171595},
url = {https://www.sciencedirect.com/science/article/pii/S0168900226003219},
author = {Yiyue Li and Huiling Li and Hui Liang and Cong Liu and Chenghan Lv and Hongbo Wang and Weiwei Xu}
}

@article{Cascella_2016,
   title={Fiber and crystals dual readout calorimeters},
   volume={31},
   ISSN={1793-656X},
   url={http://dx.doi.org/10.1142/S0217751X16440243},
   DOI={10.1142/s0217751x16440243},
   number={33},
   journal={International Journal of Modern Physics A},
   publisher={World Scientific Pub Co Pte Lt},
   author={Cascella, Michele and Franchino, Silvia and Lee, Sehwook},
   year={2016},
   month=Nov, pages={1644024} }

@techreport{Wigmans:1256562,
      author        = "Wigmans, R",
      title         = "{Dual-Readout Calorimetry for High-Quality Energy
                       Measurements}",
      institution   = "CERN",
      reportNumber  = "CERN-SPSC-2010-012, SPSC-M-771",
      address       = "Geneva",
      year          = "2010",
      url           = "https://cds.cern.ch/record/1256562",
      note          = "Supporting Document for presentation by Dr. R. Wigmans at
                       the 96th Meeting of the SPSC, April 2010",
}

@article{Ma_2022,
doi = {10.1088/1674-1137/ac3fa6},
url = {https://doi.org/10.1088/1674-1137/ac3fa6},
year = {2022},
month = {mar},
publisher = {Chinese Physical Society and the Institute of High Energy Physics of the Chinese Academy of Sciences and the Institute of Modern Physics of the Chinese Academy of Sciences and IOP Publishing Ltd},
volume = {46},
number = {3},
pages = {030001},
author = {Ma, Xin-Hua and Bi, Yu-Jiang and Cao, Zhen and Chen, Ming-Jun and Chen, Song-Zhan and Cheng, Yao-Dong and Gong, Guang-Hua and Gu, Min-Hao and He, Hui-Hai and Hou, Chao and Huang, Wen-Hao and Huang, Xing-Tao and Liu, Cheng and Shchegolev, Oleg and Sheng, Xiang-Dong and Stenkin, Yuri and Wu, Chao-Yong and Wu, Han-Rong and Wu, Sha and Xiao, Gang and Yao, Zhi-Guo and Zhang, Shou-Shan and Zhang, Yi and Zuo, Xiong},
title = {Chapter 1 LHAASO Instruments and Detector technology *},
journal = {Chinese Physics C}
}

@article{Abusleme_2022,
   title={Mass testing and characterization of 20-inch PMTs for JUNO},
   volume={82},
   ISSN={1434-6052},
   url={http://dx.doi.org/10.1140/epjc/s10052-022-11002-8},
   DOI={10.1140/epjc/s10052-022-11002-8},
   number={12},
   journal={The European Physical Journal C},
   publisher={Springer Science and Business Media LLC},
   author={Abusleme, Angel and Adam, Thomas and Ahmad, Shakeel and Ahmed, Rizwan and Aiello, Sebastiano and Akram, Muhammad and Aleem, Abid and Alexandros, Tsagkarakis and An, Fengpeng and An, Qi and Andronico, Giuseppe and Anfimov, Nikolay and Antonelli, Vito and Antoshkina, Tatiana and Asavapibhop, Burin and de André, João Pedro Athayde Marcondes and Auguste, Didier and Bai, Weidong and Balashov, Nikita and Baldini, Wander and Barresi, Andrea and Basilico, Davide and Baussan, Eric and Bellato, Marco and Bergnoli, Antonio and Birkenfeld, Thilo and Blin, Sylvie and Blum, David and Blyth, Simon and Bolshakova, Anastasia and Bongrand, Mathieu and Bordereau, Clément and Breton, Dominique and Brigatti, Augusto and Brugnera, Riccardo and Bruno, Riccardo and Budano, Antonio and Busto, Jose and Butorov, Ilya and Cabrera, Anatael and Caccianiga, Barbara and Cai, Hao and Cai, Xiao and Cai, Yanke and Cai, Zhiyan and Callegari, Riccardo and Cammi, Antonio and Campeny, Agustin and Cao, Chuanya and Cao, Guofu and Cao, Jun and Caruso, Rossella and Cerna, Cédric and Chan, Chi and Chang, Jinfan and Chang, Yun and Chen, Guoming and Chen, Pingping and Chen, Po-An and Chen, Shaomin and Chen, Xurong and Chen, Yixue and Chen, Yu and Chen, Zhiyuan and Chen, Zikang and Cheng, Jie and Cheng, Yaping and Cheng, Yu Chin and Chetverikov, Alexey and Chiesa, Davide and Chimenti, Pietro and Chukanov, Artem and Claverie, Gérard and Clementi, Catia and Clerbaux, Barbara and Colomer Molla, Marta and Conforti Di Lorenzo, Selma and Corti, Daniele and Corso, Flavio Dal and Dalager, Olivia and De La Taille, Christophe and Deng, Zhi and Deng, Ziyan and Depnering, Wilfried and Diaz, Marco and Ding, Xuefeng and Ding, Yayun and Dirgantara, Bayu and Dmitrievsky, Sergey and Dohnal, Tadeas and Dolzhikov, Dmitry and Donchenko, Georgy and Dong, Jianmeng and Doroshkevich, Evgeny and Dracos, Marcos and Druillole, Frédéric and Du, Ran and Du, Shuxian and Dusini, Stefano and Dvorak, Martin and Enqvist, Timo and Enzmann, Heike and Fabbri, Andrea and Fan, Donghua and Fan, Lei and Fang, Jian and Fang, Wenxing and Fargetta, Marco and Fedoseev, Dmitry and Fei, Zhengyong and Feng, Li-Cheng and Feng, Qichun and Ford, Richard and Fournier, Amélie and Gan, Haonan and Gao, Feng and Garfagnini, Alberto and Gavrikov, Arsenii and Giammarchi, Marco and Giudice, Nunzio and Gonchar, Maxim and Gong, Guanghua and Gong, Hui and Gornushkin, Yuri and Göttel, Alexandre and Grassi, Marco and Gromov, Vasily and Gu, Minghao and Gu, Xiaofei and Gu, Yu and Guan, Mengyun and Guan, Yuduo and Guardone, Nunzio and Guo, Cong and Guo, Jingyuan and Guo, Wanlei and Guo, Xinheng and Guo, Yuhang and Hackspacher, Paul and Hagner, Caren and Han, Ran and Han, Yang and He, Miao and He, Wei and Heinz, Tobias and Hellmuth, Patrick and Heng, Yuekun and Herrera, Rafael and Hor, YuenKeung and Hou, Shaojing and Hsiung, Yee and Hu, Bei-Zhen and Hu, Hang and Hu, Jianrun and Hu, Jun and Hu, Shouyang and Hu, Tao and Hu, Yuxiang and Hu, Zhuojun and Huang, Guihong and Huang, Hanxiong and Huang, Kaixuan and Huang, Wenhao and Huang, Xin and Huang, Xingtao and Huang, Yongbo and Hui, Jiaqi and Huo, Lei and Huo, Wenju and Huss, Cédric and Hussain, Safeer and Ioannisian, Ara and Isocrate, Roberto and Jelmini, Beatrice and Jeria, Ignacio and Ji, Xiaolu and Jia, Huihui and Jia, Junji and Jian, Siyu and Jiang, Di and Jiang, Wei and Jiang, Xiaoshan and Jing, Xiaoping and Jollet, Cécile and Joutsenvaara, Jari and Kalousis, Leonidas and Kampmann, Philipp and Kang, Li and Karaparambil, Rebin and Kazarian, Narine and Khatun, Amina and Khosonthongkee, Khanchai and Korablev, Denis and Kouzakov, Konstantin and Krasnoperov, Alexey and Kutovskiy, Nikolay and Kuusiniemi, Pasi and Lachenmaier, Tobias and Landini, Cecilia and Leblanc, Sébastien and Lebrin, Victor and Lefevre, Frederic and Lei, Ruiting and Leitner, Rupert and Leung, Jason and Li, Daozheng and Li, Demin and Li, Fei and Li, Fule and Li, Gaosong and Li, Huiling and Li, Mengzhao and Li, Min and Li, Nan and Li, Nan and Li, Qingjiang and Li, Ruhui and Li, Rui and Li, Shanfeng and Li, Tao and Li, Teng and Li, Weidong and Li, Weiguo and Li, Xiaomei and Li, Xiaonan and Li, Xinglong and Li, Yi and Li, Yichen and Li, Yufeng and Li, Zepeng and Li, Zhaohan and Li, Zhibing and Li, Ziyuan and Li, Zonghai and Liang, Hao and Liang, Hao and Liao, Jiajun and Limphirat, Ayut and Lin, Guey-Lin and Lin, Shengxin and Lin, Tao and Ling, Jiajie and Lippi, Ivano and Liu, Fang and Liu, Haidong and Liu, Haotian and Liu, Hongbang and Liu, Hongjuan and Liu, Hongtao and Liu, Hui and Liu, Jianglai and Liu, Jinchang and Liu, Min and Liu, Qian and Liu, Qin and Liu, Runxuan and Liu, Shubin and Liu, Shulin and Liu, Xiaowei and Liu, Xiwen and Liu, Yan and Liu, Yunzhe and Lokhov, Alexey and Lombardi, Paolo and Lombardo, Claudio and Loo, Kai and Lu, Chuan and Lu, Haoqi and Lu, Jingbin and Lu, Junguang and Lu, Shuxiang and Lubsandorzhiev, Bayarto and Lubsandorzhiev, Sultim and Ludhova, Livia and Lukanov, Arslan and Luo, Daibin and Luo, Fengjiao and Luo, Guang and Luo, Shu and Luo, Wuming and Luo, Xiaojie and Lyashuk, Vladimir and Ma, Bangzheng and Ma, Bing and Ma, Qiumei and Ma, Si and Ma, Xiaoyan and Ma, Xubo and Maalmi, Jihane and Mai, Jingyu and Malyshkin, Yury and Mandujano, Roberto Carlos and Mantovani, Fabio and Manzali, Francesco and Mao, Xin and Mao, Yajun and Mari, Stefano M. and Marini, Filippo and Martellini, Cristina and Martin-Chassard, Gisele and Martini, Agnese and Mayer, Matthias and Mayilyan, Davit and Mednieks, Ints and Meng, Yue and Meregaglia, Anselmo and Meroni, Emanuela and Meyhöfer, David and Mezzetto, Mauro and Miller, Jonathan and Miramonti, Lino and Montini, Paolo and Montuschi, Michele and Müller, Axel and Nastasi, Massimiliano and Naumov, Dmitry V. and Naumova, Elena and Navas-Nicolas, Diana and Nemchenok, Igor and Nguyen Thi, Minh Thuan and Ning, Feipeng and Ning, Zhe and Nunokawa, Hiroshi and Oberauer, Lothar and Ochoa-Ricoux, Juan Pedro and Olshevskiy, Alexander and Orestano, Domizia and Ortica, Fausto and Othegraven, Rainer and Paoloni, Alessandro and Parmeggiano, Sergio and Pei, Yatian and Pelicci, Luca and Pelliccia, Nicomede and Peng, Anguo and Peng, Haiping and Peng, Yu and Peng, Zhaoyuan and Perrot, Frédéric and Petitjean, Pierre-Alexandre and Petrucci, Fabrizio and Pilarczyk, Oliver and Piñeres Rico, Luis Felipe and Popov, Artyom and Poussot, Pascal and Previtali, Ezio and Qi, Fazhi and Qi, Ming and Qian, Sen and Qian, Xiaohui and Qian, Zhen and Qiao, Hao and Qin, Zhonghua and Qiu, Shoukang and Ranucci, Gioacchino and Raper, Neill and Re, Alessandra and Rebber, Henning and Rebii, Abdel and Redchuk, Mariia and Ren, Bin and Ren, Jie and Ricci, Barbara and Rifai, Mariam and Roche, Mathieu and Rodphai, Narongkiat and Romani, Aldo and Roskovec, Bedřich and Ruan, Xichao and Rybnikov, Arseniy and Sadovsky, Andrey and Saggese, Paolo and Sanfilippo, Simone and Sangka, Anut and Sawangwit, Utane and Sawatzki, Julia and Schever, Michaela and Schwab, Cédric and Schweizer, Konstantin and Selyunin, Alexandr and Serafini, Andrea and Settanta, Giulio and Settimo, Mariangela and Shao, Zhuang and Sharov, Vladislav and Shaydurova, Arina and Shi, Jingyan and Shi, Yanan and Shutov, Vitaly and Sidorenkov, Andrey and Šimkovic, Fedor and Sirignano, Chiara and Siripak, Jaruchit and Sisti, Monica and Slupecki, Maciej and Smirnov, Mikhail and Smirnov, Oleg and Sogo-Bezerra, Thiago and Sokolov, Sergey and Songwadhana, Julanan and Soonthornthum, Boonrucksar and Sotnikov, Albert and Šrámek, Ondřej and Sreethawong, Warintorn and Stahl, Achim and Stanco, Luca and Stankevich, Konstantin and Štefánik, Dušan and Steiger, Hans and Steinmann, Jochen and Sterr, Tobias and Stock, Matthias Raphael and Strati, Virginia and Studenikin, Alexander and Su, Jun and Sun, Shifeng and Sun, Xilei and Sun, Yongjie and Sun, Yongzhao and Sun, Zhengyang and Suwonjandee, Narumon and Szelezniak, Michal and Tang, Jian and Tang, Qiang and Tang, Quan and Tang, Xiao and Theisen, Eric and Tietzsch, Alexander and Tkachev, Igor and Tmej, Tomas and Torri, Marco Danilo Claudio and Treskov, Konstantin and Triossi, Andrea and Troni, Giancarlo and Trzaska, Wladyslaw and Tuve, Cristina and Ushakov, Nikita and Vedin, Vadim and Verde, Giuseppe and Vialkov, Maxim and Viaud, Benoit and Vollbrecht, Cornelius Moritz and Volpe, Cristina and von Sturm, Katharina and Vorobel, Vit and Voronin, Dmitriy and Votano, Lucia and Walker, Pablo and Wang, Caishen and Wang, Chung-Hsiang and Wang, En and Wang, Guoli and Wang, Jian and Wang, Jun and Wang, Lu and Wang, Meifen and Wang, Meng and Wang, Meng and Wang, Ruiguang and Wang, Siguang and Wang, Wei and Wang, Wei and Wang, Wenshuai and Wang, Xi and Wang, Xiangyue and Wang, Yangfu and Wang, Yaoguang and Wang, Yi and Wang, Yi and Wang, Yifang and Wang, Yuanqing and Wang, Yuman and Wang, Zhe and Wang, Zheng and Wang, Zhimin and Wang, Zongyi and Watcharangkool, Apimook and Wei, Wei and Wei, Wei and Wei, Wenlu and Wei, Yadong and Wen, Kaile and Wen, Liangjian and Wiebusch, Christopher and Wong, Steven Chan-Fai and Wonsak, Bjoern and Wu, Diru and Wu, Qun and Wu, Zhi and Wurm, Michael and Wurtz, Jacques and Wysotzki, Christian and Xi, Yufei and Xia, Dongmei and Xiao, Xiang and Xie, Xiaochuan and Xie, Yuguang and Xie, Zhangquan and Xin, Zhao and Xing, Zhizhong and Xu, Benda and Xu, Cheng and Xu, Donglian and Xu, Fanrong and Xu, Hangkun and Xu, Jilei and Xu, Jing and Xu, Meihang and Xu, Yin and Xu, Yu and Yan, Baojun and Yan, Taylor and Yan, Wenqi and Yan, Xiongbo and Yan, Yupeng and Yang, Changgen and Yang, Chengfeng and Yang, Huan and Yang, Jie and Yang, Lei and Yang, Xiaoyu and Yang, Yifan and Yang, Yifan and Yao, Haifeng and Ye, Jiaxuan and Ye, Mei and Ye, Ziping and Yermia, Frédéric and Yin, Na and You, Zhengyun and Yu, Boxiang and Yu, Chiye and Yu, Chunxu and Yu, Hongzhao and Yu, Miao and Yu, Xianghui and Yu, Zeyuan and Yu, Zezhong and Yuan, Cenxi and Yuan, Chengzhuo and Yuan, Ying and Yuan, Zhenxiong and Yue, Baobiao and Zafar, Noman and Zavadskyi, Vitalii and Zeng, Shan and Zeng, Tingxuan and Zeng, Yuda and Zhan, Liang and Zhang, Aiqiang and Zhang, Bin and Zhang, Binting and Zhang, Feiyang and Zhang, Guoqing and Zhang, Honghao and Zhang, Jialiang and Zhang, Jiawen and Zhang, Jie and Zhang, Jin and Zhang, Jingbo and Zhang, Jinnan and Zhang, Mohan and Zhang, Peng and Zhang, Qingmin and Zhang, Shiqi and Zhang, Shu and Zhang, Tao and Zhang, Xiaomei and Zhang, Xin and Zhang, Xuantong and Zhang, Xueyao and Zhang, Yinhong and Zhang, Yiyu and Zhang, Yongpeng and Zhang, Yu and Zhang, Yuanyuan and Zhang, Yumei and Zhang, Zhenyu and Zhang, Zhijian and Zhao, Fengyi and Zhao, Jie and Zhao, Rong and Zhao, Runze and Zhao, Shujun and Zheng, Dongqin and Zheng, Hua and Zheng, Yangheng and Zhong, Weirong and Zhou, Jing and Zhou, Li and Zhou, Nan and Zhou, Shun and Zhou, Tong and Zhou, Xiang and Zhu, Jiang and Zhu, Jingsen and Zhu, Kangfu and Zhu, Kejun and Zhu, Zhihang and Zhuang, Bo and Zhuang, Honglin and Zong, Liang and Zou, Jiaheng and },
   year={2022},
   month=Dec }

@article{Anfimov_2017,
doi = {10.1088/1748-0221/12/06/C06017},
url = {https://doi.org/10.1088/1748-0221/12/06/C06017},
year = {2017},
month = {jun},
publisher = {},
volume = {12},
number = {06},
pages = {C06017},
author = {Anfimov, N.},
title = {Large photocathode 20-inch PMT testing methods for the JUNO experiment},
journal = {Journal of Instrumentation}
}

@article{FUKUDA2003418,
title = {The Super-Kamiokande detector},
journal = {Nuclear Instruments and Methods in Physics Research Section A: Accelerators, Spectrometers, Detectors and Associated Equipment},
volume = {501},
number = {2},
pages = {418-462},
year = {2003},
issn = {0168-9002},
doi = {https://doi.org/10.1016/S0168-9002(03)00425-X},
url = {https://www.sciencedirect.com/science/article/pii/S016890020300425X},
author = {S. Fukuda and Y. Fukuda and T. Hayakawa and E. Ichihara and M. Ishitsuka and Y. Itow and T. Kajita and J. Kameda and K. Kaneyuki and S. Kasuga and K. Kobayashi and Y. Kobayashi and Y. Koshio and M. Miura and S. Moriyama and M. Nakahata and S. Nakayama and T. Namba and Y. Obayashi and A. Okada and M. Oketa and K. Okumura and T. Oyabu and N. Sakurai and M. Shiozawa and Y. Suzuki and Y. Takeuchi and T. Toshito and Y. Totsuka and S. Yamada and S. Desai and M. Earl and J.T. Hong and E. Kearns and M. Masuzawa and M.D. Messier and J.L. Stone and L.R. Sulak and C.W. Walter and W. Wang and K. Scholberg and T. Barszczak and D. Casper and D.W. Liu and W. Gajewski and P.G. Halverson and J. Hsu and W.R. Kropp and S. Mine and L.R. Price and F. Reines and M. Smy and H.W. Sobel and M.R. Vagins and K.S. Ganezer and W.E. Keig and R.W. Ellsworth and S. Tasaka and J.W. Flanagan and A. Kibayashi and J.G. Learned and S. Matsuno and V.J. Stenger and Y. Hayato and T. Ishii and A. Ichikawa and J. Kanzaki and T. Kobayashi and T. Maruyama and K. Nakamura and Y. Oyama and A. Sakai and M. Sakuda and O. Sasaki and S. Echigo and T. Iwashita and M. Kohama and A.T. Suzuki and M. Hasegawa and T. Inagaki and I. Kato and H. Maesaka and T. Nakaya and K. Nishikawa and S. Yamamoto and T.J. Haines and B.K. Kim and R. Sanford and R. Svoboda and E. Blaufuss and M.L. Chen and Z. Conner and J.A. Goodman and E. Guillian and G.W. Sullivan and D. Turcan and A. Habig and M. Ackerman and F. Goebel and J. Hill and C.K. Jung and T. Kato and D. Kerr and M. Malek and K. Martens and C. Mauger and C. McGrew and E. Sharkey and B. Viren and C. Yanagisawa and W. Doki and S. Inaba and K. Ito and M. Kirisawa and M. Kitaguchi and C. Mitsuda and K. Miyano and C. Saji and M. Takahata and M. Takahashi and K. Higuchi and Y. Kajiyama and A. Kusano and Y. Nagashima and K. Nitta and M. Takita and T. Yamaguchi and M. Yoshida and H.I. Kim and S.B. Kim and J. Yoo and H. Okazawa and M. Etoh and K. Fujita and Y. Gando and A. Hasegawa and T. Hasegawa and S. Hatakeyama and K. Inoue and K. Ishihara and T. Iwamoto and M. Koga and I. Nishiyama and H. Ogawa and J. Shirai and A. Suzuki and T. Takayama and F. Tsushima and M. Koshiba and Y. Ichikawa and T. Hashimoto and Y. Hatakeyama and M. Koike and T. Horiuchi and M. Nemoto and K. Nishijima and H. Takeda and H. Fujiyasu and T. Futagami and H. Ishino and Y. Kanaya and M. Morii and H. Nishihama and H. Nishimura and T. Suzuki and Y. Watanabe and D. Kielczewska and U. Golebiewska and H.G. Berns and S.B. Boyd and R.A. Doyle and J.S. George and A.L. Stachyra and L.L. Wai and R.J. Wilkes and K.K. Young and H. Kobayashi}
}

@article{ZHANG2017429,
title = {Study on the performance of electromagnetic particle detectors of LHAASO-KM2A},
journal = {Nuclear Instruments and Methods in Physics Research Section A: Accelerators, Spectrometers, Detectors and Associated Equipment},
volume = {845},
pages = {429-433},
year = {2017},
note = {Proceedings of the Vienna Conference on Instrumentation 2016},
issn = {0168-9002},
doi = {https://doi.org/10.1016/j.nima.2016.04.079},
url = {https://www.sciencedirect.com/science/article/pii/S0168900216302923},
author = {Zhongquan Zhang and Chao Hou and Zhen Cao and Jingfan Chang and Cunfeng Feng and Erlan Hanapia and Guanghua Gong and Jia Liu and Hongkui Lv and Xiangdong Sheng and Shaoru Zhang and Chengguang Zhu}
}

@article{CHO2025100021,
author = "Cho, Guk and others",
    title = "{Beam tests of the copper-based Dual-Readout calorimeter to measure electromagnetic performance for future e+e{\ensuremath{-}} colliders}",
    doi = "10.1016/j.jspc.2025.100021",
    journal = "J. Subatomic Part. Cosmol.",
    volume = "3",
    pages = "100021",
    year = "2025"
}

@article{RevModPhys.90.025002,
  title = {Dual-readout calorimetry},
  author = {Lee, Sehwook and Livan, Michele and Wigmans, Richard},
  journal = {Rev. Mod. Phys.},
  volume = {90},
  issue = {2},
  pages = {025002},
  numpages = {40},
  year = {2018},
  month = {Apr},
  publisher = {American Physical Society},
  doi = {10.1103/RevModPhys.90.025002},
  url = {https://link.aps.org/doi/10.1103/RevModPhys.90.025002}
}

@book{wigmans2000calorimetry,
  title={Calorimetry: Energy measurement in particle physics},
  author={Wigmans, Richard},
  year={2000},
  publisher={Oxford University Press}
}

@Article{jmmp8060238,
AUTHOR = {Azarhoushang, Bahman and Paknejad, Masih and Bösinger, Robert and Benner, Hans Martin},
TITLE = {The Effects of Alloy Composition and Surface Integrity on the Machinability of Austenitic Stainless Steels 304 and 304L},
JOURNAL = {Journal of Manufacturing and Materials Processing},
VOLUME = {8},
YEAR = {2024},
NUMBER = {6},
ARTICLE-NUMBER = {238},
URL = {https://www.mdpi.com/2504-4494/8/6/238},
ISSN = {2504-4494},
DOI = {10.3390/jmmp8060238}
}

@article{Yao:2024drm,
    author = "Yao, Zhao-Qian and Binosi, Daniele and Roberts, Craig D.",
    title = "{Onset of scaling violation in pion and kaon elastic electromagnetic form factors}",
    eprint = "2405.04681",
    archivePrefix = "arXiv",
    primaryClass = "hep-ph",
    reportNumber = "NJU-INP 087/24",
    doi = "10.1016/j.physletb.2024.138823",
    journal = "Phys. Lett. B",
    volume = "855",
    pages = "138823",
    year = "2024"
}

@article{Lee:2017oye,
    author = "Lee, Sehwook and Livan, Michele and Wigmans, Richard",
    title = "{On the limits of the hadronic energy resolution of calorimeters}",
    eprint = "1710.10535",
    archivePrefix = "arXiv",
    primaryClass = "physics.ins-det",
    doi = "10.1016/j.nima.2017.10.087",
    journal = "Nucl. Instrum. Meth. A",
    volume = "882",
    pages = "148--157",
    year = "2018"
}

@article{Leroy:2000mj,
    author = "Leroy, C. and Rancoita, P.",
    title = "{Physics of cascading shower generation and propagation in matter: Principles of high-energy, ultrahigh-energy and compensating calorimetry}",
    doi = "10.1088/0034-4885/63/4/202",
    journal = "Rept. Prog. Phys.",
    volume = "63",
    pages = "505--606",
    year = "2000"
}

@article{Burkert:2022hjz,
    author = "Burkert, V. D. and others",
    title = "{Precision studies of QCD in the low energy domain of the EIC}",
    eprint = "2211.15746",
    archivePrefix = "arXiv",
    primaryClass = "nucl-ex",
    doi = "10.1016/j.ppnp.2023.104032",
    journal = "Prog. Part. Nucl. Phys.",
    volume = "131",
    pages = "104032",
    year = "2023"
}

@article{Aguilar:2019teb,
    author = "Aguilar, Arlene C. and others",
    title = "{Pion and Kaon Structure at the Electron-Ion Collider}",
    eprint = "1907.08218",
    archivePrefix = "arXiv",
    primaryClass = "nucl-ex",
    reportNumber = "NJU-INP 001/19",
    doi = "10.1140/epja/i2019-12885-0",
    journal = "Eur. Phys. J. A",
    volume = "55",
    number = "10",
    pages = "190",
    year = "2019"
}

@article{Wigmans:2018fua,
    author = "Wigmans, Richard",
    title = "{New developments in calorimetric particle detection}",
    eprint = "1807.03853",
    archivePrefix = "arXiv",
    primaryClass = "physics.ins-det",
    doi = "10.1016/j.ppnp.2018.07.003",
    journal = "Prog. Part. Nucl. Phys.",
    volume = "103",
    pages = "109--161",
    year = "2018"
}

@article{Fabjan:2003aq,
    author = "Fabjan, C. W. and Gianotti, F.",
    title = "{Calorimetry for particle physics}",
    reportNumber = "CERN-EP-2003-075",
    doi = "10.1103/RevModPhys.75.1243",
    journal = "Rev. Mod. Phys.",
    volume = "75",
    pages = "1243--1286",
    year = "2003"
}

@article{Harter:2023efw,
    author = "Harter, A. and Weinert, M. and Knafla, L. and R{\'e}gis, J. -M. and Esmaylzadeh, A. and Ley, M. and Jolie, J.",
    title = "{Systematic investigation of time walk and time resolution characteristics of CAEN digitizers V1730 and V1751 for application to fast-timing lifetime measurement}",
    eprint = "2303.07946",
    archivePrefix = "arXiv",
    primaryClass = "physics.ins-det",
    doi = "10.1016/j.nima.2023.168356",
    journal = "Nucl. Instrum. Meth. A",
    volume = "1053",
    pages = "168356",
    year = "2023"
}

@article{Milton:2025zdc,
    author = {Milton, Ryan and Paul, Sebouh J. and Schmookler, Barak
              and Arratia, Miguel and Karande, Piyush and Angerami, Aaron
              and Torales Acosta, Fernando and Nachman, Benjamin},
    title = {Design and simulation of a {SiPM}-on-tile {ZDC} for the future
             {EIC}, and its performance with graph neural networks},
    journal = {Nucl. Instrum. Meth. A},
    volume = {1079},
    pages = {170613},
    year = {2025},
    doi = {10.1016/j.nima.2025.170613},
    eprint = {2406.12877},
    archivePrefix = {arXiv},
    primaryClass = {physics.ins-det}
}

@article{Paul:2025lambda,
    author = {Paul, Sebouh J. and Milton, Ryan and Morán, Sebastián
              and Schmookler, Barak and Arratia, Miguel},
    title = {Feasibility Study of Measuring
             {$\Lambda^{0}\to n\pi^{0}$} Using a High-Granularity
             Zero-Degree Calorimeter at the Future Electron-Ion Collider},
    journal = {Phys. Rev. D},
    volume = {111},
    pages = {092013},
    year = {2025},
    doi = {10.1103/q7w9-sbsc},
    eprint = {2412.12346},
    archivePrefix = {arXiv},
    primaryClass = {nucl-ex}
}

@phdthesis{Zhang:2017LHAASO,
  author = {Zhongquan Zhang},
  title  = {Performance Study and Optimization of the Electromagnetic Particle Detector for {LHAASO-KM2A}},
  school = {Shandong University},
  year   = {2017},
  note   = {Ph.D. thesis, in Chinese}
}






\end{document}